%% file: main.tex
\documentclass[usenatbib,usenatbib]{mnras}

\DeclareRobustCommand{\VAN}[3]{#2}
\let\VANthebibliography\thebibliography
\def\thebibliography{\DeclareRobustCommand{\VAN}[3]{##3}\VANthebibliography}

\usepackage{graphicx}	
\usepackage{amsmath}	
\usepackage{amssymb}	
\usepackage{newtxtext,newtxmath}
\usepackage{physics}
\usepackage{xspace}
\usepackage{psfrag}
\usepackage{epsfig}
\usepackage{caption}
\usepackage{pdfpages}
\usepackage[T1]{fontenc}

\newcommand{\be}{\begin{equation}}
\newcommand{\e}{\end{equation}}
\newcommand{\bear}{\begin{eqnarray}}
\newcommand{\ear}{\end{eqnarray}}

\newcommand{\comment}[1]{}

\def\xi{x^{i}_{\ion{H}{i}}\,}
\def\xh1{x_{\ion{H}{i}\,}}

\def\x{x_{\ion{H}{i}}}
\def\Ph1{P_{\ion{H}{i}}}
\def\Bh1{B_{\ion{H}{i}}}
\def\eh1{\eta_{\ion{H}{i}}}

\def\aap{AAP}
\def\apj{ApJ}
\def\aj{AJ}
\def\apjs{ApJS}
\def\apjl{ApJL}
\def\mnras{MNRAS}
\def\prd{PRD}

\def\HI{\ion{H}{i}\xspace}

\def\P{\hat{P}}

\def\mp{\, {\rm Mpc}^{-1}}
\def\tb{\bar{{\delta T_{\rm b}}}}

\def\mk{\, {\rm mK}}

\def\ve{{\em very early}\,}
\def\ee{{\em early}\,}
\def\m{{\em middle}\,}
\def\l{{\em late}\,}
\def\vl{{\em very late}\,}

\def\ss{{\em small}\,}
\def\ii{{\em intermediate}\,}
\def\ll{{\em large}\,}

\newcommand{\TB}{\delta T_{\rm b}}
\newcommand{\MSUN}{{\rm M}_{\odot}}

\newcommand{\XHII}{x_{\rm HII}}
\newcommand{\TS}{T_{\rm S}}
\newcommand{\TK}{T_{\rm K}}
\newcommand{\TCMB}{T_{\gamma}}
\newcommand{\lya}{\rm {Ly{\alpha}}}

\newcommand{\Omegam}{\Omega_{\rm m}}

\usepackage[normalem]{ulem}

\def\RG{\textcolor{cyan}}

\title[CD 21-cm bispectrum]{Redshifted 21-cm bispectrum II: Impact of the spin temperature fluctuations and redshift space distortions on the signal from the Cosmic Dawn}

\author[Kamran et al.]{Mohd Kamran$^{1}$\thanks{E-mail: kamranmohd080@gmail.com, phd1801121002@iiti.ac.in},
Raghunath Ghara$^{2,3,4}$,
Suman Majumdar$^{1,5}$,
Rajesh Mondal$^4$,\newauthor Garrelt Mellema$^{4}$, Somnath Bharadwaj$^{6}$, Jonathan R. Pritchard$^{5}$, Ilian T. Iliev$^{7}$
\\
$^{1}$Department of Astronomy, Astrophysics and Space Engineering, Indian Institute of Technology Indore, Simrol, Indore 453552, India\\
$^{2}$ARCO (Astrophysics Research Center), Department of Natural Sciences, The Open University of Israel, 1 University Road, PO Box 808, Ra'anana 4353701, Israel \\
$^{3}$Department of Physics, Technion, Haifa 32000, Israel\\
$^{4}$The Oskar Klein Centre, Department of Astronomy, Stockholm University, AlbaNova, SE-10691 Stockholm, Sweden\\
$^{5}$Department of Physics, Blackett Laboratory, Imperial College, London SW7 2AZ, U. K.\\
$^{6}$Department of Physics and Centre for Theoretical Studies, Indian Institute of Technology Kharagpur, Kharagpur - 721 302, India\\ $^{7}$Astronomy Centre, Department of Physics and Astronomy, Pevensey II Building, University of Sussex, Brighton BN1 9QH, UK\\
}
\date{Accepted XXX. Received YYY; in original form ZZZ}

\begin{document}
\label{firstpage}
\pagerange{\pageref{firstpage}--\pageref{lastpage}}
\maketitle


\input{abstract}
\input{introduction}

\input{bispec_theory}

\input{simulation}
\input{result}
\input{summary}
\input{acknow}

\section*{Data Availability}
The simulated data underlying this article will be shared on reasonable request to the corresponding author.





\bibliographystyle{mnras}
\bibliography{reference,mybib} 




\label{lastpage}
\end{document}

%% file: abstract.tex
\begin{abstract}
We present a study of the 21-cm signal bispectrum (which quantifies the non-Gaussianity in the signal) from the Cosmic Dawn (CD). For our analysis, we have simulated the 21-cm signal using radiative transfer code {\sc grizzly}, while considering two types of sources (mini-QSOs and HMXBs) for Ly$\alpha$ coupling and the X-ray heating of the IGM. Using this simulated signal, we have, for the first time, estimated the CD 21-cm bispectra for all unique $k$-triangles and for a range of $k$ modes. We observe that the redshift evolution of the bispectra magnitude and sign follow a generic trend for both source models. However, the redshifts at which the bispectra magnitude reach their maximum and minimum values and show their sign reversal depends on the source model. When the Ly$\alpha$ coupling and the X-ray heating of the IGM occur simultaneously, we observe two consecutive sign reversals in the bispectra for small $k$-triangles (irrespective of the source models). One arising at the beginning of the IGM heating and the other at the end of Ly$\alpha$ coupling saturation. This feature can be used in principle to constrain the CD history and/or to identify the specific CD scenarios. We also quantify the impact of the spin temperature ($\TS$) fluctuations on the bispectra. We find that $\TS$ fluctuations have maximum impact on the bispectra magnitude for small $k$-triangles and at the stage when Ly$\alpha$ coupling reaches saturation. Furthermore, we are also the first to quantify the impact of redshift space distortions (RSD), on the CD bispectra. We find that the impact of RSD on the CD 21-cm bispectra is significant ($> 20\%$) and the level depends on the stages of the CD and the $k$-triangles for which the bispectra are being estimated.

\end{abstract}  


%% file: introduction.tex
\section{Introduction}
\label{sec:intro}
The appearance of the first stars and galaxies in the Universe is often referred to as the Cosmic Dawn (CD). In the subsequent epoch, known as Epoch of Reionization (EoR), the neutral hydrogen ($\HI$) in the Inter-Galactic Medium (IGM) was ionized by the radiation from those early sources. The CD-EoR epoch is one of the least understood chapters from the evolutionary history of our Universe. Observations of the redshifted $\HI$ 21-cm signal produced by the spin-flip transition of the electron in the ground state of hydrogen, will provide a wealth of information about the thermal and ionization states of the IGM during the CD/EoR as well as about the properties of the first sources in the early Universe (see e.g. \citealt{barkana01, furlanetto06, pritchard12}). 

A number of radio interferometers such as the GMRT\footnote{\url{http://www.gmrt.ncra.tifr.res.in}} \citep{Paciga:2013}, LOFAR\footnote{\url{http://www.lofar.org/}} \citep{mertens20}, MWA\footnote{\url{http://www.mwatelescope.org/}} \citep{barry19}, PAPER \citep{kolopanis19} and HERA\footnote{\url{https://reionization.org/}} \citep{2017PASP..129d5001D} have dedicated considerable amounts of observing time to detect the CD-EoR 21-cm signal. Due to their limited sensitivity, these interferometers can only hope to detect the redshifted 21-cm signal statistically using the spherically averaged power spectrum as an estimator. The future Square Kilometre Array (SKA)\footnote{\url{http://www.skatelescope.org/}} will however have sufficient sensitivity to also produce tomographic images of the 21-cm signal \citep{2015aska.confE..10M, koopmans15, ghara16}. 

There is a considerable observational effort underway to detect the 21-cm power spectrum (see e.g. \citealt{ali08, ghosh12, mertens20} etc.). So far, only weak upper limits on the expected EoR 21-cm power spectrum have been obtained \citep{mertens20, barry19, li19, kolopanis19, trott20}. Apart from power spectrum, several authors have proposed other statistics to detect the 21-cm signal such as variance \citep[e.g.][] {iliev08, patil14, watkinson14, watkinson15}, multi-frequency angular power spectrum \citep[e.g.][] {datta07a,mondal18,mondal19,mondal20a} etc. 

The power spectrum is nothing but the variance at different length scales and therefore, is able to capture many crucial features of the signal (see e.g., \citealt{bharadwaj04,bharadwaj05, barkana05, datta07a,datta14, mesinger07, lidz08, choudhury09b, mao12, majumdar13,majumdar14, majumdar16, jensen13, 2019MNRAS.tmp.1183R}). It can capture the complete statistical properties of a Gaussian random field, for which all higher-order statistics do not contain any additional information. However, the 21-cm signal is highly non-Gaussian in nature \citep{bharadwaj05a, mellema06, mondal15}. The non-Gaussianity in the EoR 21-cm signal evolves as reionization proceeds \citep{mondal16, mondal17, shaw19, shaw20}. Thus power spectrum will not be able to characterize this signal completely. Some of the one-point statistics such as skewness and kurtosis can capture this non-Gaussianity (see e.g., \citealt{harker09, watkinson14, watkinson15, shimabukuro15a, kubota16, 2020arXiv201103558R} etc). However, these can only probe the non-Gaussian features at a single length scale. To probe the non-Gaussian features at different length scales, one requires some robust higher-order statistics such as the bispectrum. 

The bispectrum, the Fourier transform of the three point correlation function, can capture the non-Gaussian component of the signal in different Fourier modes. The non-Gaussian features in the EoR 21-cm signal were first studied using the bispectrum by \citet{bharadwaj05} who employed an analytical model of spherically ionized regions. They reported that the sign of the bispectrum can have negative values. \citet{majumdar18} estimated the EoR 21-cm bispectrum using a suite of semi-numerical simulations, focusing on some specific $k$-triangles (equilateral, isosceles etc.). They confirmed that the bispectrum displays sign changes and showed that the interplay between neutral fraction and matter density fluctuations decides the sign of the bispectrum. A negative bispectrum indicates that the non-Gaussianity is arising due to fluctuations in the neutral fraction whereas a positive bispectrum occurs when the non-Gaussianity is arising due to the matter density fluctuations. \citet{hutter19} independently produced similar results. Furthermore, in the context of the CD, \citet{watkinson19} have calculated the bispectrum (with a different normalization) for some specific $k$-triangles by using a fully-numerical simulation. They showed that bispectrum is driven by the interplay between the size and shape of the heated regions and their distributions. They also found that irrespective of the type of the X-ray heating sources, the bispectrum magnitude is largest for equilateral $k$-triangle configurations. \citet{shimabukuro16} presented another independent study of the CD-EoR 21-cm bispectrum for some specific types of $k$-triangles. However, their estimator was unable to capture the sign of the bispectrum. 

The peculiar velocity of the gas modifies the cosmological redshift of the 21-cm signal and makes it anisotropic along the LoS. This LoS anisotropy in the signal is known as redshift space distortions (RSD). In the context of the CD-EoR 21-cm signal, their effect was first calculated by \citet{bharadwaj04} using analytical models for the 21-cm observable. They showed that the RSD has a large impact on the amplitude and shape of the 21-cm power spectrum. Subsequent works studied the RSD using semi-numerical and fully-numerical simulations \citet{mao12, majumdar13, majumdar14, majumdar16, jensen13, ghara15a,  fialkov15, mondal18}. 

All previous studies of the CD-EoR 21-cm bispectrum were restricted to certain specific triangle configurations and neglected the RSD effect. For a comprehensive description of the non-Gaussianity of a field through bispectrum, all possible closed unique $k$-triangles should be considered. In a recent study, \citealt{bharadwaj20} have shown a way to parametrize the real and redshift space bispectrum for all unique $k$-triangles. Further, \citet{saxena20, majumdar20} have studied the bispectrum for all unique $k$-triangles using simulations of the EoR 21-cm signal. \citet{saxena20} demonstrated how different dark matter models impact the EoR bispectrum through their impact on the structure formation. In \citet{majumdar20} (hereafter Paper-I), we have also considered the impact of the RSD on the EoR 21-cm signal bispectrum. We showed that all possible unique $k$-triangles capture the non-Gaussianity arising due to two competing sources i.e., matter density fluctuations and the neutral hydrogen fluctuations. We have also shown that the squeezed and linear $k$-triangles capture the largest magnitude of the bispectrum, and they are highly sensitive to the non-Gaussianity arising due to the matter density fluctuations at small scales. We further showed that the RSD impacts both the magnitude and the sign of the bispectrum irrespective of the stages of the EoR and size and shape of the $k$-triangles.


\begin{figure*}
  \includegraphics[width=0.47\textwidth,angle=0]{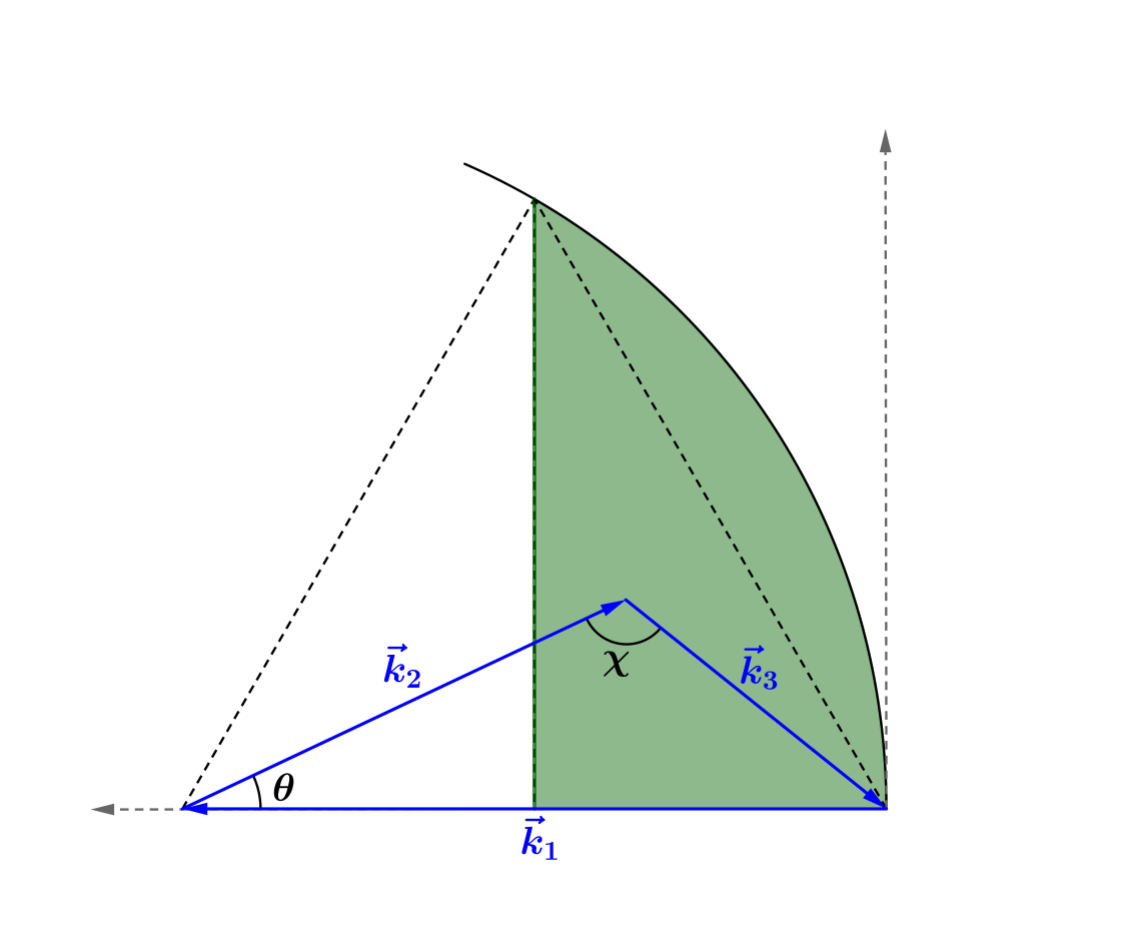}
  \includegraphics[width=0.47\textwidth,angle=0]{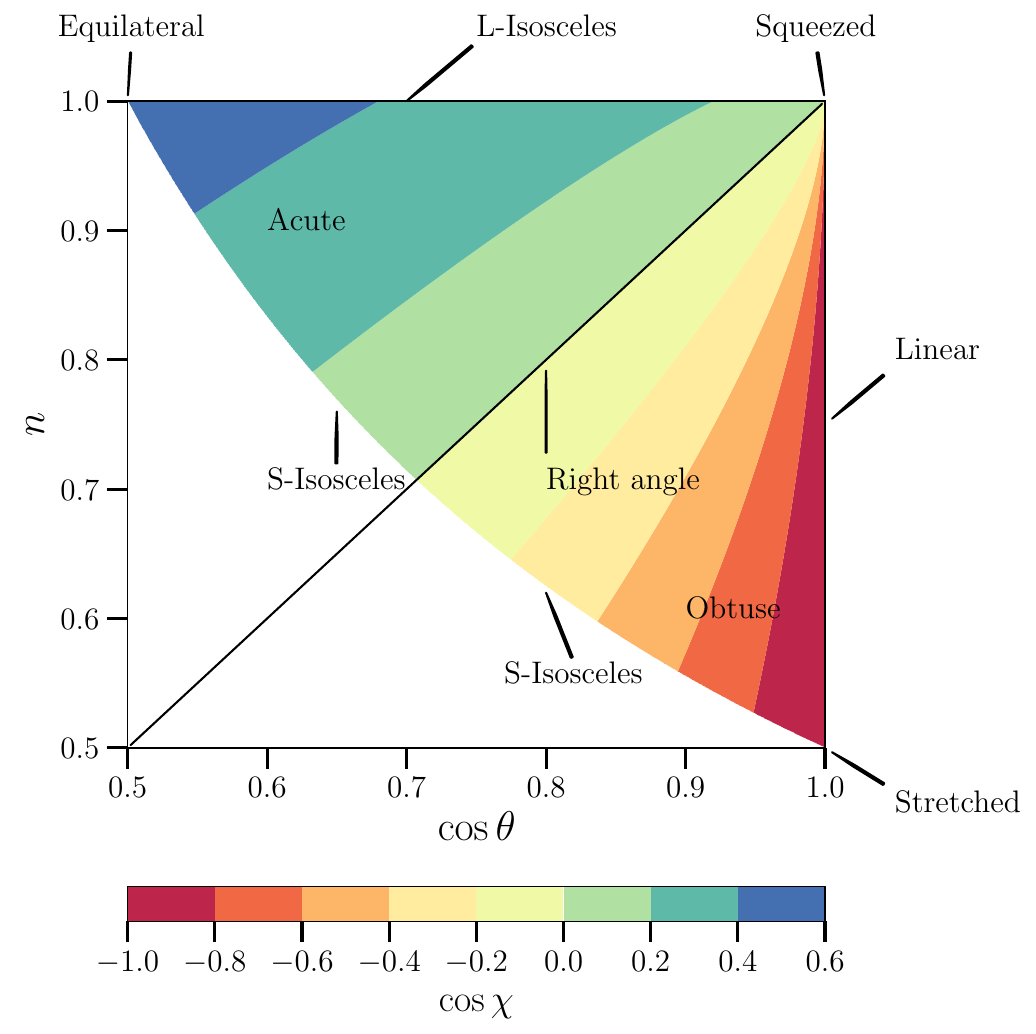}
  \caption{\textbf{Left panel:} It shows the definition of unique triangles in ${\bf k}$-space. The shaded region shows the all possible positions of the point of intersection between $\textbf{k}_2$ and $\textbf{k}_3$ vectors while equations \eqref{eq:k_relation}, \eqref{eq:n} and \eqref{eq:costheta} are satisfied. \textbf{Right panel:} It shows the variation of $\cos{\chi}$ with $n$ and $\cos{\theta}$ for all unique $k$-triangles. The region $n\cos{\theta} \geq 0.5$ is bounded by the L-isosceles, S-isosceles and linear $k$-triangles. The corner points represents equilateral, squeezed and stretched limit of $k$-triangles.}
  \label{figures:triangle_uni}
\end{figure*}
This article aims to quantify the non-Gaussianity in the CD 21-cm signal through the bispectrum while considering all possible unique $k$-triangles. We do not include the reionization part (discussed in details in Paper I) of the cosmic history here. This is the first study of the redshift space 21-cm bispectrum from the CD considering all the possible unique $k$-triangles. 

The structure of the paper is as follows: In Section \ref{sec:bispec_theory} we present the formalism for estimating the bispectrum from the simulated signal. Section \ref{sec:sim} describes the simulations used to generate the 21-cm maps. The Section \ref{sec:results} we discuss our analysis of the bispectra for different source models. Finally in Section \ref{sec:summary} we summarise our results.


In this paper, we have used the cosmological parameters $h = 0.7$, $\Omega_{\mathrm{m}} = 0.27$, $\Omega_{\Lambda} = 0.73$, $\Omega_{\mathrm{b}} = 0.044$ consistent with the \textit{WMAP} results \citep{hinshaw13} and within the error bars consistent with the \textit{Planck} results \citep{planck14}.

\comment{The observable quantity for the redshifted $\HI$ 21-cm signal is the differential brightness temperature which corresponds to intensity of the 21-cm radiation with respect to intensity of Cosmic Microwave Background (CMB) along the line of sight (LoS) and can be expressed as  \citep[see e.g.][]{madau1997, furlanetto06, pritchard12, mao12, mellema13} 
\begin{multline}
\delta T_{\rm b}(\textbf{x}, z_{\rm cos}) = 27  \x(\textbf{x}, z_{\rm cos}) \frac{\left(1+\delta_{\rm b}(\textbf{x},z_{\rm cos})\right)}{\Big|1+\frac{1+z_{\rm cos}}{H(z_{\rm cos})} \frac{dv_{||}(\textbf{x})}{dr_{||}}\Big|} \Bigg( \frac{\Omega_{\rm b} h^2}{0.023} \Bigg)  \\
\times \! \Bigg( \frac{0.15}{\Omega_{\rm m} h^2} \frac{1+z_{\rm cos}}{10} \Bigg)^{1/2}\Bigg(1-\frac{T_{\rm CMB}}{T_{\rm S}} \Bigg) \Bigg(\frac{1-Y_{\rm p}}{1-0.248}\Bigg) ~\rm mK.
\label{eq:tb_real}
\end{multline}

Here $\x$ and $\delta_{\rm b}$ denote the hydrogen neutral fraction and baryonic density contrast respectively at a location $\textbf{x} (= r_{\nu} \hat{n})$ with comoving distance $r_{\nu}$ along the line of sight $\hat{n}$ and at redshift $z_{\rm cos}$. Further, $\frac{1+z_{\rm cos}}{H(z_{\rm cos})} \frac{dv_{||}(\textbf{x})}{dr_{||}}$ is the gradient of peculiar velocity along the line of sight normalized by conformal Hubble parameter $H(z_{\rm cos})/(1+z_{\rm cos})$. 

Equation \ref{eq:tb_real} is dependent on the cosmological parameters: The Hubble parameter ${H(z_{\rm cos})}$ = 100h, baryon density parameter ($\Omega_{\rm b}$) and matter density parameter ($\Omega_{\rm m}$) where, $\Omega_{\rm i} = \rho_{\rm i}/\rho_{\rm c}$ and $\rho_{\rm c}$ is the critical density for the spatially flat universe, $Y_{\rm p}$ is primordial helium fraction by mass.
}

\comment{Throughout this paper the differential brightness temperature involved in our analysis depends on the fluctuations in the neutral fraction of hydrogen ($\delta_{\x}$), density contrast of the baryonic matter ($\delta_{\rm b}$), fluctuations in the spin-temperature ($T_{\rm S}$) and peculiar velocity of the gas along the Line-of-Sight (LoS). The spin temperature provides us a measure of relative population of the neutral hydrogen over the two energy states associated with the 21-cm hyperfine transition. Just after the formation of the first radiation sources, they produce a large amount of the Lyman-$\alpha$  photons and couple the $T_{\rm S}$ with the gas temperature ($T_{\rm k}$) through Wouthuysen–Field effect (\citealt{wouth52, field58}). However, brightness temperature fluctuations are efficiently sourced by the fluctuations in Lyman-$\alpha$  flux through $T_{\rm S}$ (\citealt{barkana05b, pritchard06}). Once the Lyman-$\alpha$  fluctuations get saturated, fluctuations in the $T_{\rm k}$ pump the brightness temperature fluctuations through $T_{\rm S}$. 
}




%% file: bispec_theory.tex
\section{Bispectrum estimation for all unique triangle configurations}
\label{sec:bispec_theory}
In this section, we present the formalism for estimating the bispectra for all unique $k$-triangles in Fourier space.

\subsection{The bispectrum and the different triangle configurations}
\label{sec:trig_config}

We adopt the bispectrum estimator defined in \cite{majumdar18} to compute the bispectrum from the simulated data for the $i^{\rm th}$ triangle configuration bin as
\begin{equation}
  \hat{B}_{i}({\bf k}_1, {\bf k}_2, {\bf k}_3) = \frac{1}{N_{{\rm tri}}V} \sum_{
[{\bf k}_1 +{\bf k}_2 +{\bf k}_3 =0] \in i }\Delta_{21}(\textbf{k}_1) \Delta_{21}(\textbf{k}_2) \Delta_{21}(\textbf{k}_3) \, ,
\label{eq:bispec_est}
\end{equation}
where $\Delta_{21}(\textbf{k})$ is the Fourier transform of the differential brightness temperature of the 21-cm field, $N_{\rm tri}$ is the number of statistically independent samples of closed triangles associated with the $i^{\rm th}$ triangle configuration bin, $V$ is the observational volume. The actual calulcation of the bispectrum is performed using the algorithm of \citet{majumdar18} and \citet{majumdar20}. To reduce the computation time for bispectrum estimation this algorithm introduced two additional constraining equations,
\begin{equation}
    \frac{k_2}{k_1} = n
    \label{ratio}
\end{equation}

\begin{equation}
    \cos{\theta} = -\frac{{\mathbf k}_1\cdot {\mathbf k}_2}{k_1 k_2}.
    \label{cos_theta}
\end{equation}
Where $k_1$ and $k_2$ are the magnitudes of the vectors ${\mathbf k}_1$ and ${\mathbf k}_2$ respectively. For a detailed discussion on this algorithm we refer the reader to \citet{majumdar18} and \citet{majumdar20}.

\comment{
 \begin{equation}
    \cos{\chi} = \frac{n^2+[1+n^2-2n\cos{\theta}]-1}{2n \sqrt{1+n^2-2n\cos{\theta}}}
    \label{eq:cos_chi}
\end{equation}
}
 
 \subsection{The unique triangle configurations in the triangle parameter space}
 \label{sec:uni_trig_config}

For a triangle of a specific size, determined by the magnitude of the $k$ modes involved, its shape can be uniquely specified in the $n-\cos\theta$ space by imposing the conditions prescribed in \citet{bharadwaj20}, i.e.,
 \begin{eqnarray}
&&{k}_1 \geq {k}_2 \geq {k}_3
\label{eq:k_relation}\\
&&0.5 \leq n \leq 1.0
\label{eq:n}\\
&&0.5 \leq \cos\theta \leq 1.0.
\label{eq:costheta}
\end{eqnarray}

The above conditions [equations \eqref{eq:k_relation}, \eqref{eq:n} and \eqref{eq:costheta}] imply that, the unique triangles occupy the region in the $n-\cos{\theta}$ space where $n\cos{\theta} \geq{0.5}$ as shown in the right panel of Figure \ref{figures:triangle_uni}. The shaded region in the left panel of this figure shows the all possible positions of the point of intersection between $\textbf{k}_2$ and $\textbf{k}_3$ vectors while equations \eqref{eq:k_relation}, \eqref{eq:n} and \eqref{eq:costheta} are satisfied. This figure (left panel) also shows all possible unique $k$-triangles represented by the cosine of the angle subtended by the vertex of triangle facing the $k_1$ arm i.e., $\cos{\chi}$. For a detailed classification of unique $k$-triangles we refer interested reader to section 2.2 of \citet{majumdar20}.

To estimate the spherically averaged bispectrum from a simulated data cube, we divide the entire range of $n$ as well as $\cos{\theta}$ with step sizes $\Delta n=0.05$ and $\Delta\cos{\theta}=0.01$. We further bin the entire $k_1$-range ($k_{\rm min} = 2\pi$/[box size] $\approx 0.01 \mp$ , $k_{\rm max} = 2\pi$/2[grid spacing]$ \approx 2.64 \mp$) into $15$ logarithmic bins. We label each bin by the value of $k_1$ as the magnitude of the $k_1$ bin determines the size of the $k$-triangle in our formalism. 

\comment{
The unique $k$-triangles with different shapes in $n-\cos{\theta}$ space (right panel of Figure \ref{figures:triangle_uni}) are listed here:
\begin{itemize}
    \item L-isosceles are triangles having $n = 1.0$ and $0.5 \lesssim \cos{\theta} \lesssim 1.0$ i.e. $0 \lesssim \cos{\chi} \lesssim 0.5$. These triangles have  $k_1=k_2$.
    \item The condition $n\cos{\theta} = 0.5$ arc in the parameter space defines S-isosceles triangles. These $k$-triangles have $k_2 = k_3$.
    \item The intersection point of the L and S isosceles triangles represents equilateral triangle i.e. $\cos{\theta}=0.5$ and $n=1.0$ (i.e. $\cos{\chi}=0.5$).
    \item Triangles with $\cos{\theta}\rightarrow1.0\,$ and $0.5 \lesssim n \lesssim 1.0$ are linear triangles. At these limits all three $k$s become collinear (i.e. $\cos{\chi}\rightarrow-1.0$).
    \item Triangles with $n=\cos{\theta}$ are right angle triangles (i.e. $\cos{\chi}=0$).
    \item The intersection point of L-isosceles, linear and right angle triangles represents squeezed triangles where $\cos{\theta}=n=1.0$ (i.e. $\cos\chi=0$). For squeezed triangles the smallest arm $k_3 \rightarrow 0$.
    \item  The intersection point of S-isosceles and linear triangles defines stretched triangles where $\cos\theta=1.0$ and $n=0.5$ (i.e. $\cos{\chi}=-1.0$).
    \item Triangles with $\cos{\theta}<n$ are acute angle triangles (i.e. $\cos{\chi>0}$).
    \item Triangles with $\cos{\theta}>n$ are obtuse angle triangles (i.e. $\cos{\chi<0}$).
\end{itemize}
}


\comment{
\section{The reconstructed bispectrum based on the linear theory of redshift space distortions}
\label{sec:B_rec}
To interpret the simulated redshift space 21-cm bispectra ($B^{\rm s}$) we introduce a model based on the linear approximation of the RSD. This model treats the impact of the RSD on the real space signal bispectra ($B^{\rm r}$)\footnote{The superscripts ``r'' and ``s'' in this article represent the quantity in real and redshift spaces respectively.} as an additional set of correction terms. 
We follow the formalism proposed by \citet{kaiser87, hamilton98} and \citet{mao12} to model the  redshift space differential brightness temperature in Fourier space as:
\begin{equation}
   \Delta_{21}^{\rm {s}}(\textbf{k}) =  \Delta_{21}^{\rm r}(\textbf{k})+\Delta^{\rm r}_{21,\,\rho_{\rm H}}(\textbf{k}) \mu_{\textbf{k}}^2 
   \label{Tb_rsd_modified}
\end{equation}
where, 
\begin{equation}
    \Delta^{\rm r}_{21,\,\rho_{\rm H}}(\textbf{k})  = \Delta^{\rm r}_{21}(\textbf{k}) \Delta^{\rm r}_{\rho_{\rm H}}(\textbf{k})\, .
\end{equation}

The $B^{\rm r}$ depends only on the size and shape of the triangle. However, the $B^{\rm s}$ in addition will depend also on the orientation of the $k$-triangle with respect to the line-of-sight (LoS). Under the plane parallel approximation the orientation dependence goes as $\mu_1^2$, $\mu_2^2$ and $\mu_3^2$ where, $\mu_i = \hat{x}\cdot \textbf{k}_i/k_i$ with $i = 1\,, 2\,, 3\,$ (the cosine of angles between the three arms of the $k$-triangle $\textbf{k}_1$, $\textbf{k}_2$, $\textbf{k}_3$ with LoS direction $\hat{x}$). Further, for a detailed analysis of the entire LoS anisotropy introduced by the RSD, one should in principle represent the redshift space bispectrum in terms of its different angular multipole moments in the orthonormal basis of spherical harmonics (${Y}^m_{\ell}$) (\citealt{scoccimarro1999, bharadwaj20}). The different angular multipole moments of redshift space bispectrum  ($\bar{B}^m_{\ell}$) under linear theory of RSD can be expressed as,
\begin{multline}
   \bar{B}^m_{\ell}(k_1, n, \cos{\theta}) =  \delta_{\ell,0}   B^{\rm r}_{\Delta_{21}, \Delta_{21}, \Delta_{21}} + \Big(
    [\overline{\mu_3^2}]^m_{\ell}  B^{\rm r}_{\Delta_{21}, \Delta_{21}, \Delta_{21,\rho_{\rm H}}}\\+
    [\overline{\mu_2^2}]^m_{\ell}   B^{\rm r}_{\Delta_{21}, \Delta_{21,\rho_{\rm H}}, \Delta_{21}} +
    [\overline{\mu_1^2}]^m_{\ell} B^{\rm r}_{\Delta_{21,\rho_{\rm H}}, \Delta_{21}, \Delta_{21}} \Big)\\ + \Big(
    [\overline{\mu_1^2 \mu_2^2}]^m_{\ell}   B^{\rm r}_{\Delta_{21,\rho_{\rm H}}, \Delta_{21,\rho_{\rm H}}, \Delta_{21}}+
    [\overline{\mu_1^2 \mu_3^2}]^m_{\ell}   B^{\rm r}_{\Delta_{21,\rho_{\rm H}}, \Delta_{21}, \Delta_{21,\rho_{\rm H}}} \\+
    [\overline{\mu_2^2 \mu_3^2}]^m_{\ell}  B^{\rm r}_{\Delta_{21}, \Delta_{21,\rho_{\rm H}}, \Delta_{21,\rho_{\rm H}}} \Big) +  [\overline{\mu_1^2 \mu_2^2 \mu_3^2}]^m_{\ell} B^{\rm r}_{\Delta_{21,\rho_{\rm H}}, \Delta_{21,\rho_{\rm H}}, \Delta_{21,\rho_{\rm H}}} \,.
   \label{eq:b_qlin_multipole}
\end{multline}

\RG{Somewhere specify the $\overline{\mu}$ quantities. I guess these are the average values.  }

The monopole moment ($\ell=0$ and $m=0$) of equation \eqref{eq:b_qlin_multipole} is the spherically averaged redshift space bispectrum estimated by averaging different orientations of the triangles with respect to the LoS and is expressed by
\begin{multline}
   B^{\rm s}({\bf k_1},{\bf k_2},{\bf k_3}) = B^{\rm r}_{\Delta_{21}, \Delta_{21}, \Delta_{21}} + \Big(
    [\overline{\mu_3^2}]^0_{0}  B^{\rm r}_{\Delta_{21}, \Delta_{21}, \Delta_{21,\rho_{\rm H}}}\\+
    [\overline{\mu_2^2}]^0_{0}   B^{\rm r}_{\Delta_{21}, \Delta_{21,\rho_{\rm H}}, \Delta_{21}} +
    [\overline{\mu_1^2}]^0_{0} B^{\rm r}_{\Delta_{21,\rho_{\rm H}}, \Delta_{21}, \Delta_{21}} \Big)\\ + \Big(
    [\overline{\mu_1^2 \mu_2^2}]^0_{0}   B^{\rm r}_{\Delta_{21,\rho_{\rm H}}, \Delta_{21,\rho_{\rm H}}, \Delta_{21}}+
    [\overline{\mu_1^2 \mu_3^2}]^0_{0}   B^{\rm r}_{\Delta_{21,\rho_{\rm H}}, \Delta_{21}, \Delta_{21,\rho_{\rm H}}} \\+
    [\overline{\mu_2^2 \mu_3^2}]^0_{0}  B^{\rm r}_{\Delta_{21}, \Delta_{21,\rho_{\rm H}}, \Delta_{21,\rho_{\rm H}}} \Big) +  [\overline{\mu_1^2 \mu_2^2 \mu_3^2}]^0_{0} B^{\rm r}_{\Delta_{21,\rho_{\rm H}}, \Delta_{21,\rho_{\rm H}}, \Delta_{21,\rho_{\rm H}}}\,.
   \label{eq:b_qlin}
\end{multline}
For a detailed discussion of the monopole moment of the coefficients in Equation \eqref{eq:b_qlin_multipole}, we refer the interested reader to section 3 of \citet{majumdar20}.

In the absence of RSD only the first term, the real space bispectrum, will remain in Equation \eqref{eq:b_qlin}. The other seven terms can therefore be designated as the redshift space correction (RC) terms. We divide these RC terms into three groups based on the power of the $\mu$ factor in each term. These groups are called $\mu^2$-RC, $\mu^4$-RC and $\mu^6$-RC.  \sout{\RG{The first two groups contain three terms and the last group contains one term.}} \RG{Each of the first two groups contains three terms while the last group contains one term.} A full description of the approach can be found in \citet{majumdar20}.
}


\comment{The bispectrum of the $\HI$ 21-cm field is defined as the Fourier counterpart of three point correlation function of this field in real space and is given by

 \begin{equation}
    \big \langle  \Delta_{21}(\textbf{k}_1) \Delta_{21}(\textbf{k}_2) \Delta_{21}(\textbf{k}_3)  \big \rangle = V\delta_{\textbf{k}_1+\textbf{k}_2+\textbf{k}_3,\:0}\: B(\textbf{k}_1,\textbf{k}_2,\textbf{k}_3) 
    \label{bispectrum}
 \end{equation}
Where, $\Delta_{21}(\textbf{k})$ is the Fourier representation of $\delta T_{\rm b}(\textbf{x})$, $V$ is observational volume and $\delta_{\textbf{k}_1+\textbf{k}_2+\textbf{k}_3,\:0}$ is the Kronecker delta function. Our bispectrum estimator will be non-zero iff three $\textbf{k}$'s involved in the definition forms a closed triangle (see Figure \ref{figures:triangle_uni}) and given given by the condition:

 \begin{equation}
    \textbf{k}_1+\textbf{k}_2+\textbf{k}_3=0
    \label{closed_tri}
 \end{equation}
This can be labelled as $k$-triangle.
}

%% file: simulation.tex
\section{Simulating the 21-cm signal from cosmic dawn}
\label{sec:sim}
We produce 21-cm differential brightness temperature ($\TB$) maps using the {\sc grizzly} \citep{ghara15a} simulations. This simulation approximates the impact of each source using spherically symmetric solutions of the radiative transfer equation. It is an independent implementation of the {\sc bears} algorithm \citep{Thom08,Thom09,Thom11}. We refer the interested readers to \citet{ghara15a, ghara18, 2020MNRAS.493.4728G} for further details. 

\subsection{Cosmological simulations}
\label{sec:nbody}
{\sc grizzly} uses gridded versions of cosmological density and velocity fields and halo catalogues at different redshift as inputs to produce $\TB$ maps at those redshifts.
The density and velocity fields used in this study were obtained as part of the PRACE\footnote{Partnership for Advanced Computing in Europe: \url{http://www.prace-ri.eu/}} project {\sc PRACE4LOFAR}. This dark-matter only $N$-body simulation was performed using the {\sc cubep$^3$m} code \citep{Harnois12}. The details of this simulation are as follows: ($i$) simulation volume:  $(500\,h^{-1})^3$~ comoving Mpc$^3$. ($ii$) number of particles: $6912^3$, ($iii$) mass of each particle: $4.05\times 10^7 ~\MSUN$. We use gridded versions of the density and velocity fields with a resolution of $600^3$. The minimum mass of the halos resolved in this simulation is $\approx 10^9$~M$_\odot$ which means that these halos consist of at least $\approx 25$ particles.

\subsection{Source properties}
\label{sec:sources}

\begin{table}
 \begin{center}
 \begin{tabular}{||c c c||} 
 \hline
Source Models\\ and Scenarios & Model-A &  Model-B\\ [0.5ex] 
 \hline\hline
 Source type & mini-QSO &  HMXB \\ 
 \hline
 Spectral index ($\alpha$) & 1.5 & 0.2 \\
 \hline
 Spin Temperature ($\TS$) & Self consistent & Self consistent \\
 \hline
\end{tabular}
\caption{Different scenarios considered in this study. We choose the star formation efficiency $f_\star=0.03$, minimum mass of dark matter halos that contain sources $M_{\rm halo, min} = 2\times10^9 ~\MSUN$, escape fraction of UV photons $f_{\rm esc}=0.1$ and the rate of emission of the X-ray photons per stellar mass $\dot{N}_X = 4\times 10^{42} ~\rm s^{-1} ~\MSUN^{-1}$ for both the scenarios.}
\label{table1}
\end{center}
\end{table}

\begin{figure}
\includegraphics[width=\columnwidth,angle=0]{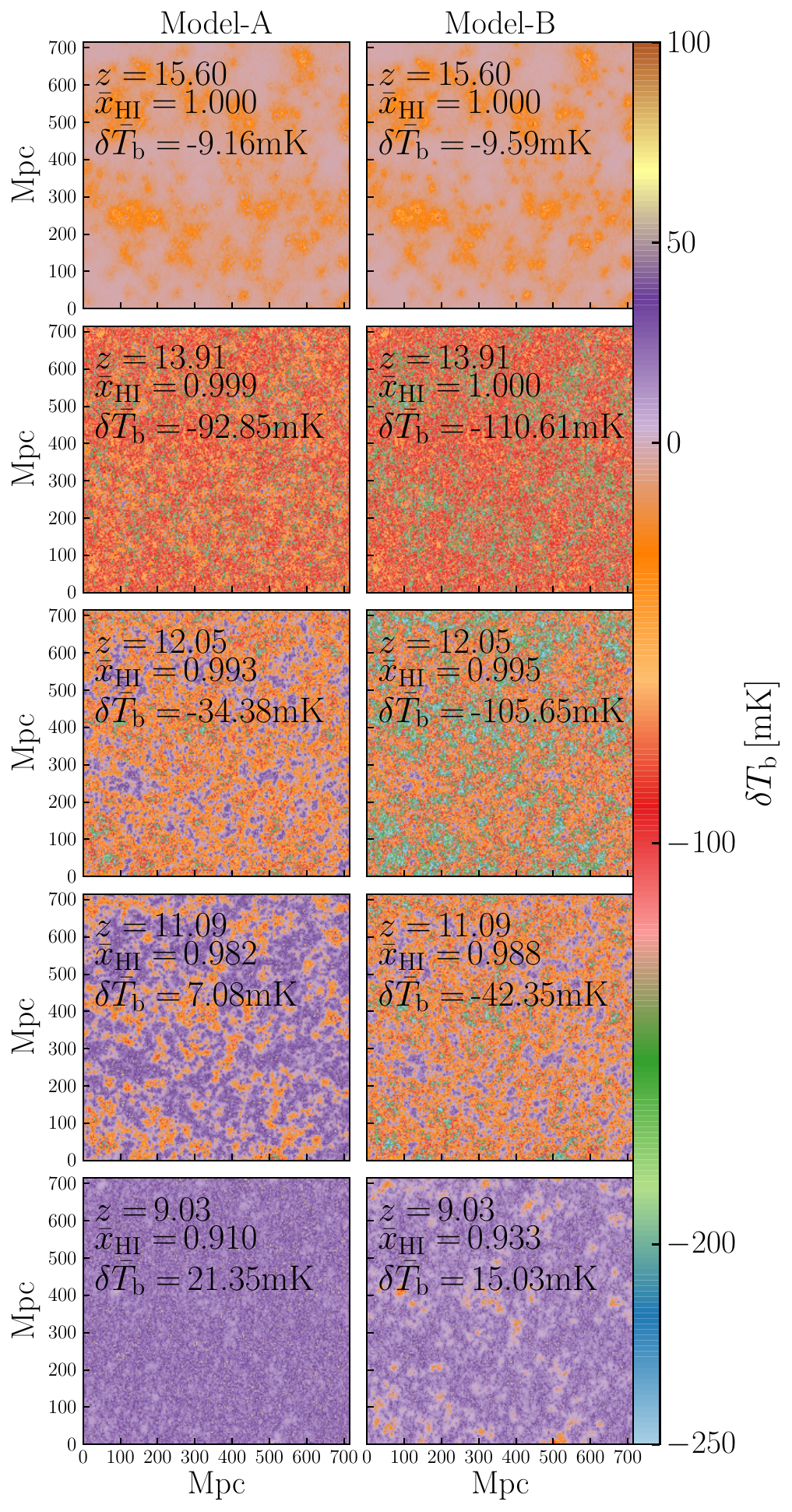}
\caption {Slices of the brightness temperature showing the redshift evolution of the 21-cm signals for Model-A (Left panels) and Model-B (right panels).}
\label{fig:tb_map}
\end{figure}

In this simulation we assume that all the dark matter halos with mass larger than the cut-off mass $M_{\rm halo, min}$ host star forming galaxies, sources of hard X-ray spectra such as high-mass X-ray binaries (HMXBs) and sources of soft X-ray spectra such as mini-quasars (mini-QSOs) etc. We further assume that most \HI ionizing photons are produced by the stars in galaxies. The stellar content of a galaxy is assumed to be proportional to the mass of the dark matter halo that hosts the galaxy. Thus, the stellar mass inside a halo of mass $M_{\rm halo}$ is  $M_\star=f_\star \left(\frac{\Omega_{\mathrm b}}{\Omegam}\right) M_{\rm halo}$ where  $f_\star$ is the star formation efficiency. We choose $f_\star = 0.03$ \citep{2015ApJ...799...32B, 2016MNRAS.460..417S} throughout this paper.

We use the publicly available stellar population synthesis code {\sc pegase2}\footnote{\tt http://www2.iap.fr/pegase/} \citep{Fioc97} to generate the SED of the galaxies assuming a stellar metallicity of $0.01$, a mass range of $1$ to $100$ $\MSUN$ and a Salpeter initial stellar mass function. The stellar luminosity in this source model scales with the stellar mass of the galaxies.

In addition to the star formation efficiency and the stellar luminosity, the actual number of \HI ionizing photons that enter the IGM from the galaxies also depends on the escape fraction ($f_{\rm esc}$) of those ionizing photons. We assume $f_{\rm esc}=0.1$ for the ionizing photons produced inside the galaxies.


Apart from the ionizing photons produced by the stellar sources, we also consider contributions from the X-ray sources such as mini-QSO and HMXBs. Unlike the stellar UV photons, X-rays can travel over long distances through the neutral IGM before they are absorbed.  We model the X-ray part of the SED as a power-low of energy $I_X(E) \propto E^{-\alpha}$, where $\alpha$ is the spectral index of the X-ray source. In our study, we consider two different $\alpha$ values 1.5 and 0.2 which roughly represents mini-QSOs \citep{2003AJ....125..433V, 2017MNRAS.467.3590G, 2017A&A...608A..51M} and HMXBs \citep{2012MNRAS.419.2095M, 2019MNRAS.487.2785I}. The former we refer to as Model-A, the latter as Model-B. We assume that the rate of emission of the X-ray photons per stellar mass is $\dot{N}_X = 4\times 10^{42} ~\rm s^{-1} ~\MSUN^{-1}$ for both types of X-ray sources. This is consistent with the observations of HMXBs in local star forming
galaxies in the 0.5-8 keV band \citep[see e.g.,][]{2012MNRAS.419.2095M}. Our X-ray band spans from 100 eV to 10 keV, while the UV band spans between 13.6-100 eV. The normalization of the X-ray band used in the 1D radiative transfer is done using the X-ray band spanning to 10 keV. Note that the hard X-ray photons with energy $\gtrsim 2$ keV will remain unabsorbed in our simulation volume due to their long mean free path.

\subsection{21-cm signal simulation}
\label{gr_21}
Given the SED of a source, {\sc grizzly} first generates a large number of one-dimensional profiles of the ionization fraction ($\XHII$) and gas temperature ($\TK$) for various combinations of the source parameter values at different redshifts. The appropriate 1D profiles are then applied to each dark matter halo in the 3D simulation volume to generate the $\XHII$ and $\TK$ cubes at the redshifts of interest. The algorithm applies a photon/energy conserving correction to overlapping regions. Although this method is based on several approximations such as isotropic radiative transfer, overlap correction, etc., it still perform well when compared to a full radiative transfer simulation with the C$^2$-{\sc ray} code \citep{mellema06} \citep[see the details in][]{ghara18}. 

{\sc grizzly} also takes into account an inhomogeneous $\lya$ background assuming that the $\lya$ flux scales as $1/r^2$ with radial distance $r$ from the source. The $\lya$ photon number density determines the strength of coupling between the spin-temperature and the gas temperature \citep[Wouthuysen-Field effect,][]{wouth52, hirata2006lya}. Although  $\lya$ photons can also heat the gas in the IGM \citep{2020MNRAS.492..634G, mittal20}, their impact is usually much less than the X-rays and thus we do not include $\lya$ heating in this paper. We also do not include the systematic sources of uncertainties in models of thermal histories of the IGM (e.g. underlying cosmology, the choice of halo mass function etc.), which significantly affects the Thomson scattering optical depth and global 21-cm signal amplitude \citep{jordan20}). An important point to note here is that in these simulations we have considered the Ly$\alpha$ coupling and the X-ray heating of the IGM to be running simultaneously (i.e. $\TS$ is calculated in a self consistent manner) from very early stages of the CD.

The redshift space distortions are implemented in our simulation using the following principle. If the redshift $z$ is caused only due to the Hubble expansion, then the redshift space position \textbf{\textit{s}} of the source of the signal will be same as its comoving real space position \textbf{\textit{r}}. However, due to the gas peculiar velocities along the LoS ($v_{||}$), the signal coming from the location \textbf{\textit{r}} will appear to be coming from a location \textbf{\textit{s}}, to the present day observer, following
\begin{equation}
    \textbf{\textit{s}} = \textbf{\textit{r}} + \frac{1+z}{H(z)}v_{||}\hat{r}
\end{equation}
where, $\hat{r}$ is the unit vector along the LoS.

We generate $\TB$ cubes at 22 redshift values in the range 9 to 20 using the $\XHII$, $\TK$ and 
the $\lya$ coupling coefficient cubes. Next  we apply the effect of the RSD \citet{mao12, majumdar13, majumdar16, jensen13, ghara15a, ghara15b} to these $\TB$ cubes. For this we use the cell movement method (or Mesh-to-Mesh (MM)-RRM scheme) of \citet{mao12}.

\subsection{Scenarios}
\label{sec:scen}
We consider two different source scenarios in this study. These two scenarios only differ in the 
spectral index of the X-ray sources. In Model-A we use $\alpha=1.5$, the value typical for mini-QSOs. In Model-B we use $\alpha=0.2$, typical for HMXBs. All other parameters
such as the minimum mass of halos used $M_{\rm halo, min}$, the escape fraction for UV photons $f_{\rm esc}$, etc., are identical. Table \ref{table1} lists all relevant source and simulation parameter values used here. 

\begin{figure}
\includegraphics[width=\columnwidth,angle=0]{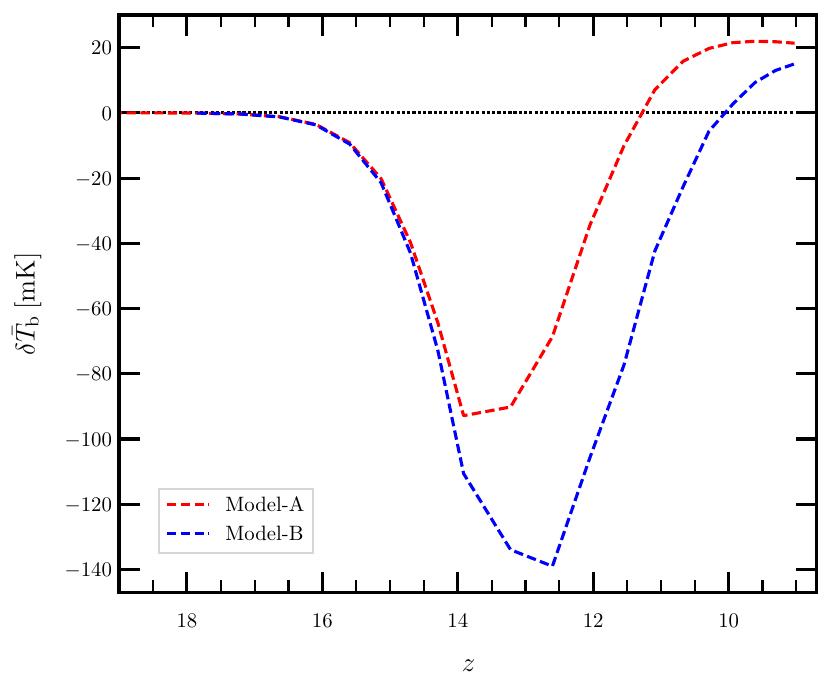}
\caption{Evolution of global 21-cm signal with redshift for mimi-QSO (red curve) and HMXB (blue curve) sources.}
\label{fig:tb_z}
\end{figure}

%% file: result.tex
\section{Results}
\label{sec:results}

\begin{figure*}
  \includegraphics[width=1\textwidth,angle=0]{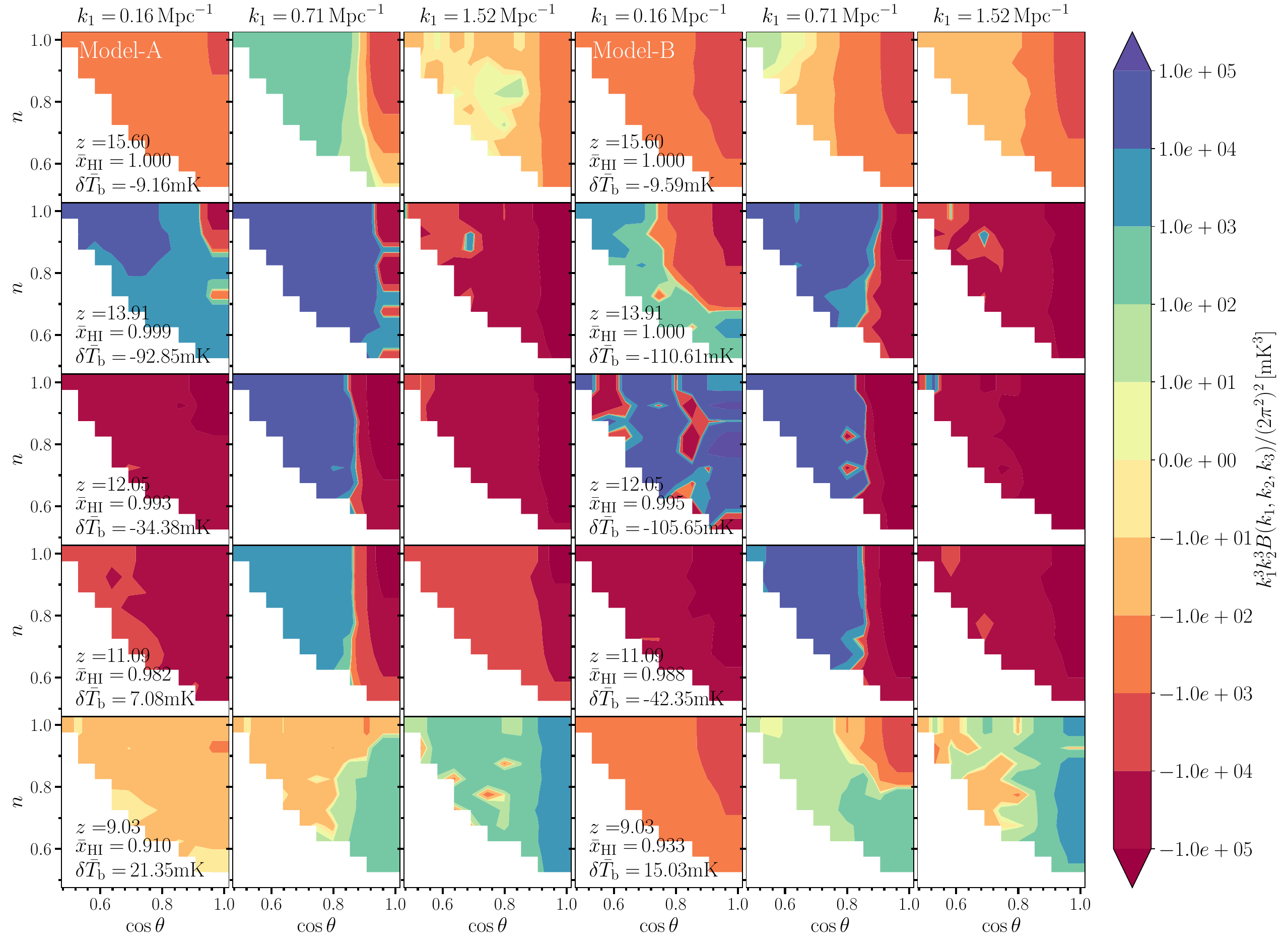}
  \caption{The redshift space bispectra for all unique triangle configurations for Model-A (left three columns) and Model-B (right three columns) at five different stages of the CD and for three different $k_1$ modes.}
  \label{fig:b_model_AB}
\end{figure*}


In this section, we present the analysis of the 21-cm bispectrum from the Cosmic Dawn for the two different Ly$\alpha$ coupling and X-ray heating scenarios we have introduced in Section \ref{sec:sources}. Figure \ref{fig:tb_map} shows the slices of $\TB$ maps for Model-A and Model-B, while the evolution of the 21-cm mean brightness temperature for two cases are shown in Figure \ref{fig:tb_z}. In this paper, we are interested in the impact of the spin temperature fluctuations on the signal statistics of these two models, for which they show significant variation in the mean brightness temperature evolution (Figure \ref{fig:tb_z}). The effects of $\lya$ coupling are the same for both models. However, the effects of X-ray heating are different. The heating of the gas happens at a later time in Model-B as compared to Model-A. This is because the source spectrum used in Model-B contains less number of soft X-ray photons as compared to a steeper source spectrum used in Model-A. The soft X-ray photons get absorbed easily due to their smaller mean free path as compared to the hard X-ray photons.

Among the two source models that we have simulated, we choose Model-A (i.e.  mini-QSOs with halo mass $\geq 2 \times 10^9\, {\rm M_{\odot}}$, see Table \ref{table1} for more details) to be our fiducial source model. Following the convention of \citet{majumdar20} and \citet{saxena20}, we demonstrate our results in terms of the spherically averaged normalized bispectrum defined as $\big[ k_1^3 k_2^3 B(k_1, k_2, k_3)/(2\pi^2)^2\big]$. Throughout this paper we show the 21-cm bispectra estimated from the redshift space data, unless otherwise stated. To designate different stages of the CD we use the following convention:  {\em very early} ($z \sim 16$), {\em early} ($z\sim 14$), {\em middle} ($z\sim 12$), {\em late} ($z\sim 11$) and {\em very late} ($z\sim 9$). Further, we label the triangles with $k_1$ modes $\approx 0.2, 0.7 \, \rm and\, 1.5 \mp$ as {\em small}, {\em intermediate} and {\em large} $k_1$-triangles respectively{\footnote{Due to our limited simulation volume, the triangle bins having $k_1$ smaller than $0.16 \mp$ will be affected significantly by the sample variance.}}. The $\delta{T}_{\rm b}$ slices as shown in Figure \ref{fig:tb_map} represent those five different stages of the IGM in this study.


\subsection{Bispectrum for the fiducial source model}
\label{sec:B_CD_RSD}
Here we first try to understand the general trends in the bispectra magnitude and sign for the fiducial model as a function of redshift and magnitude of the $k_1$ mode. Below we divide our discussion for the fiducial model into two segments: evolution of the bispectrum magnitude and evolution of the bispectrum sign.


\begin{figure*}
  \includegraphics[width=1\textwidth,angle=0]{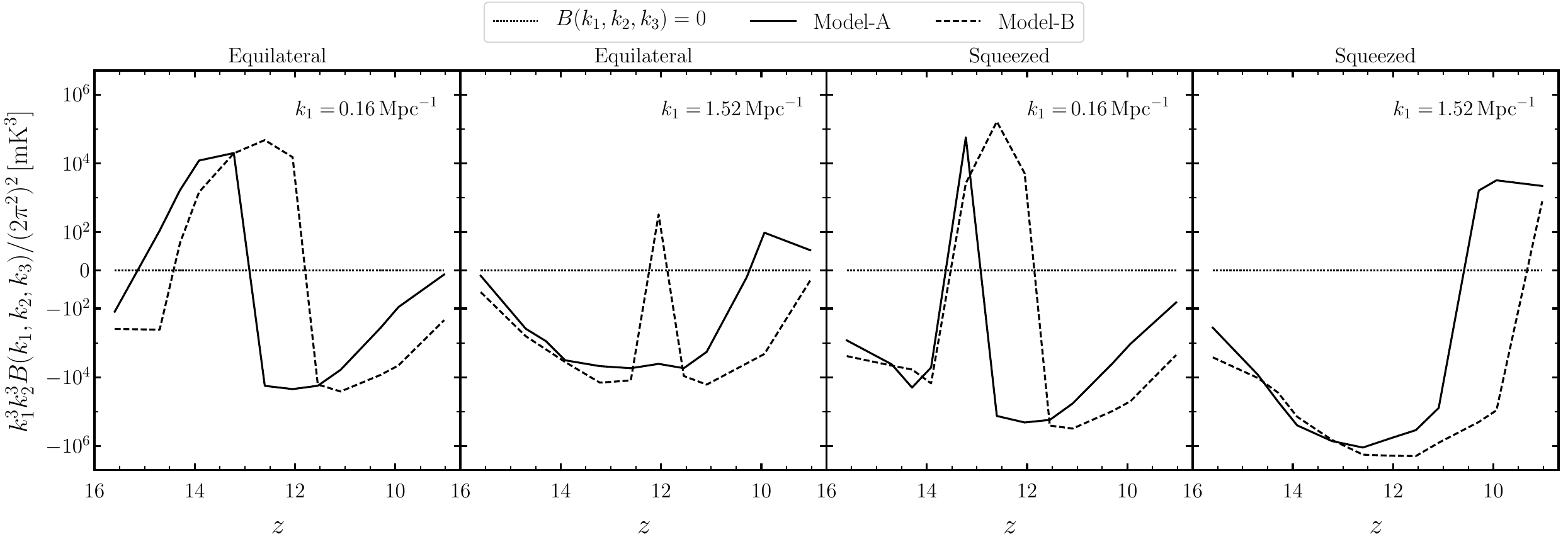}
  \caption{Variation of the bispectrum with redshift for different source models for two types of triangles and two different $k_1$ modes.}
  \label{fig:BS_z}
\end{figure*}

\subsubsection{Evolution of the bispectrum magnitude}
The first three columns in Figure \ref{fig:b_model_AB} show the normalized spherically averaged bispectra for our fiducial model at the five different stages of the CD (varying from top to bottom) and for the three different $k_1$ modes (varying from left to right) as defined in Section \ref{sec:results}. A quick inspection of Figure \ref{fig:tb_z} (red dashed curve for the fiducial model) reveals that the mean 21-cm signal is in absorption during most of the CD and it turns into an emission signal around $z \approx 11.5$. This implies that for our fiducial model the first three rows of Figure \ref{fig:b_model_AB} show the bispectrum when the signal is in absorption and the bottom two rows show the bispectrum when the signal is in emission.

We first discuss the variation of the magnitude of the bispectra with $z$. For ease of understanding, in addition to Figure \ref{fig:b_model_AB}, we plot the bispectra for equilateral and squeezed triangles as a function of redshift in Figure \ref{fig:BS_z}. A visual inspection of the Figures \ref{fig:b_model_AB} and \ref{fig:BS_z} reveals that the magnitude of the bispectrum is lowest during the \ve and the \vl stages of the CD. This magnitude is much larger (more than $\sim 2$  orders of magnitude) during the \ee, \m and \l stages of the CD compared to the beginning and the end of the CD. This is due to the fact that during the intermediate stage of the CD the fluctuations in the signal becomes maximum. 
The reason for the low magnitude of the bispectrum at the \ve and \vl stages of the CD can be understood in the following manner. The fluctuations in the 21-cm signal during the CD can potentially have many constituents. Among them are the fluctuations in the matter density field, fluctuations in the hydrogen neutral fractions and the fluctuations in the spin temperature. Among these three, the fluctuations in the neutral fraction stays relatively low in amplitude during the entire duration of the CD as only up to $\sim 10\%$ of neutral hydrogen ionizes by the end of the CD (around the \vl stage). The magnitude of the fluctuations in the matter density is even lower during this period when compared to the other constituents of the signal fluctuations. Additionally, the fluctuations in the $T_S$ field is relatively small during the \ve stages because there are very few absorption and emission regions around this time and they are also very small in their sizes. This leads to a relatively small magnitude of bispectrum at the \ve stages of the CD.


Around the \vl stages $T_{\rm S}$ becomes much larger compared to the CMB temperature ($\TCMB$), and the signal enters into the EoR era with a significant ionization (less than $10\%$). At this stage, the components responsible for 21-cm signal fluctuations are the matter density and hydrogen neutral fraction fluctuations. The fluctuations in these components around \vl stages are also relatively small. This makes the bispectrum magnitude smaller during the \vl stages. This is a direct evidence of the fact that the CD 21-cm signal non-Gaussianity quantified by the bispectrum is relatively higher than the EoR 21-cm signal non-Gaussianity. 

An important point to note here is regarding the magnitude of the bispectra during the \l stages. Around this stage, although the global signal is of emission in nature with a relatively small mean brightness temperature (i.e., $\tb \sim 7 \mk$), the bispectrum magnitude is still relatively large. This is due to the additional high fluctuations introduced in the field by some left over absorption regions (see panels in the fourth row of Figure \ref{fig:tb_map}).

Next, we discuss the variation of the bispectrum magnitude with $k_1$. Around \ve stage, the magnitude of the bispectrum is maximum at the limit and vicinity of squeezed ($n=1, \cos{\theta}=1$) $k_1$-triangles. This largest magnitude patch in the $n$--$\cos{\theta}$ space expands its area around the linear $k$-triangles with increasing values of $k_1$. The bispectrum magnitude for the rest of $k_1$-triangles at this stage is smaller by $\sim 1$--$2$ orders of magnitude. 
Around \ee stage and for \ss $k_1$ mode, the maximum bispectrum magnitude is observed for the equilateral, squeezed, and similar class of $k$-triangles. For these $k$-triangles with larger $k_1$ values, the bispectrum magnitude more or less remains the same, but for the rest of the $k$-triangles in this class, it increases with an increase in $k_1$ magnitude. Further, the bispectrum magnitude around \m stage does not show a significant variation with $k_1$. Next, around the \l stage, the bispectrum magnitude decreases with increase in $k_1$ values except for the squeezed and the linear $k$-triangles.
Lastly, around \vl stage, the bispectrum magnitude increases with $k_1$, and its largest variations are seen for the linear $k$-triangles. 
Hence, for most of the time during the CD, bispectrum will have maximum magnitude in the limit and the vicinity of squeezed and linear $k$-triangles.
\begin{figure*}
  \includegraphics[width=1\textwidth,angle=0]{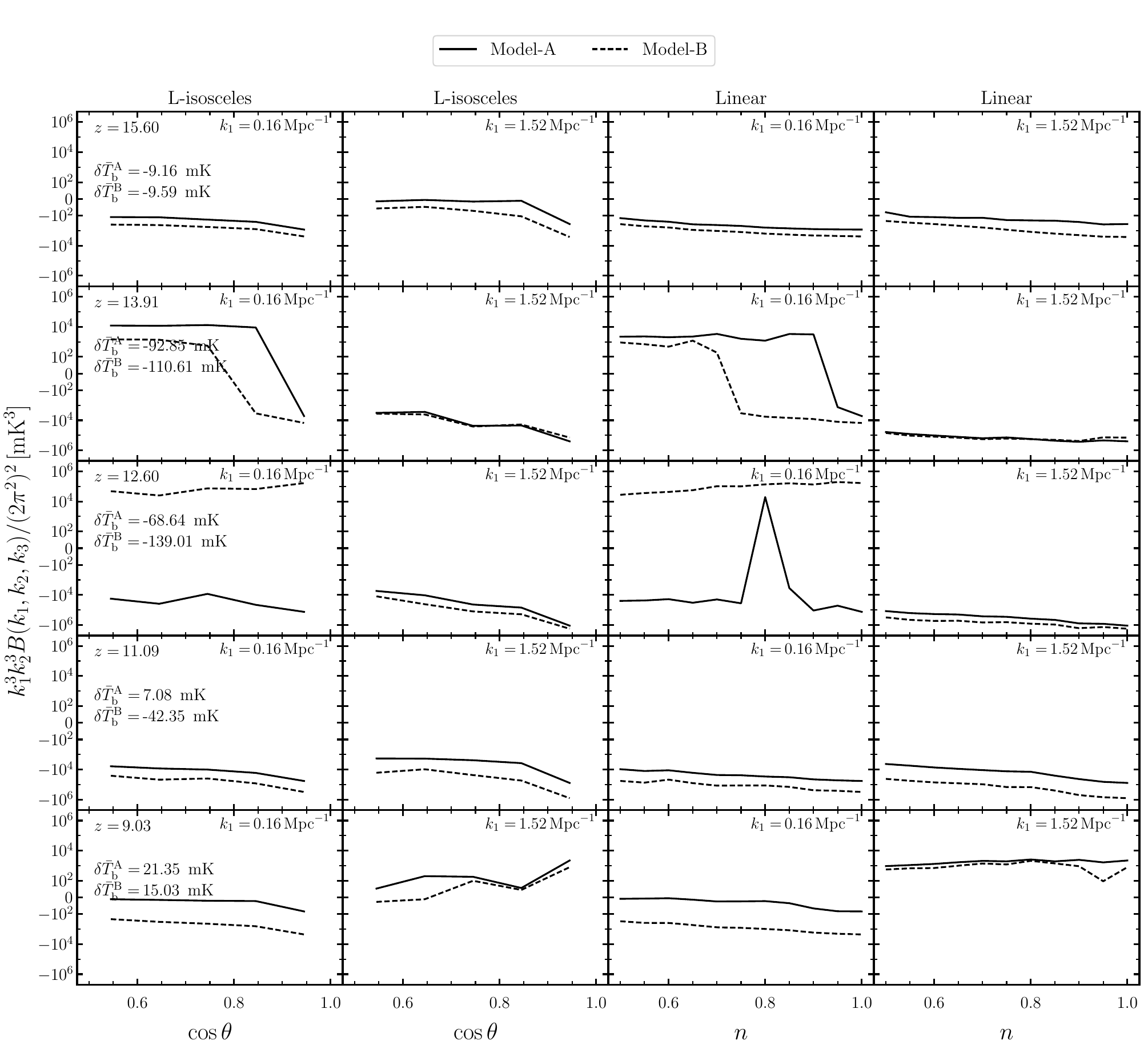}
  \caption{The redshift space bispectra for L-isosceles ($\cos{\theta},\, n = 1$) and linear ($n,\, \cos{\theta} = 1$) $k$-triangles for Model-A (solid lines) and Model-B (dashed line) at five different stages of CD and for \ss and \ll $k_1$ modes.}
  \label{fig:line_bispec_all_model}
\end{figure*}
\begin{figure*}
  \includegraphics[width=1\textwidth,angle=0]{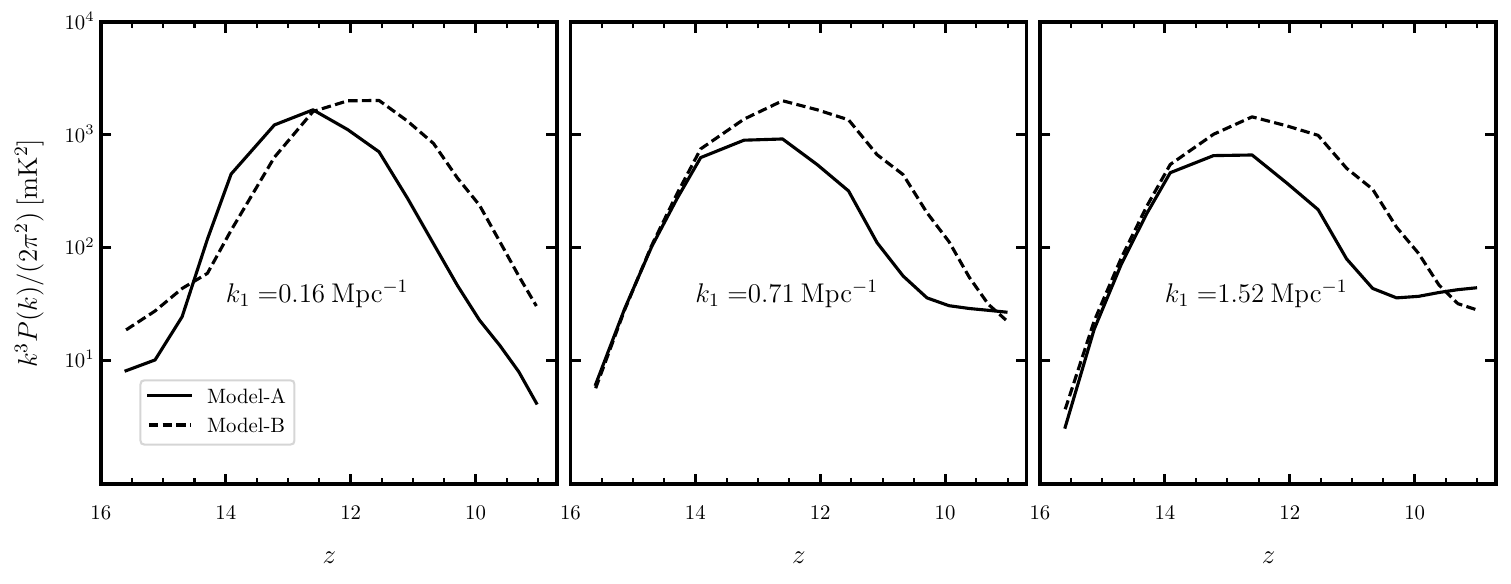}
  \caption{Variation of the power spectrum with redshift for different source models for three different $k_1$ modes.}
  \label{fig:pk}
\end{figure*}
\subsubsection{Evolution of the bispectrum sign}
We next focus on the sign of the bispectrum which has been identified as an important feature of this statistic along with its magnitude in several earlier works \citep{majumdar18,majumdar20,hutter19, watkinson19}. In Figure \ref{fig:b_model_AB} for Model-A we observe that around \ve stage $B(k_1 \approx 0.2 \mp$) is negative for the entire $n-\cos\theta$ space. This is due to the fact that at \ve stage the size and number density of Ly$\alpha$ coupled regions are very large. Whereas, the size and number density of the heated regions are comparatively very low (see the top left panel of Figure \ref{fig:tb_map}). The bispectrum can have a negative sign due to either of the following two reasons: if $\Delta_{21}$s for all three $k$-modes have negative sign or the $\Delta_{21}$ for only one $k$ mode has negative sign. Due to the dominance of the absorption regions during this period (Figure \ref{fig:tb_map}), it is most likely that the first cause is satisfied in this case. For the \ll $k_1$-triangles in the entire $n$--$\cos{\theta}$ space the bispectra remain negative until the \l stage. This behaviour is observed for the period when $T_{\rm S}$ fluctuations dominate over the other two constituent fields. A detailed discussion on this behaviour is further given in Section \ref{sec:impact_of_Ts}. 

We also observe that there are certain regions in the $n-\cos{\theta}$ space where the bispectra are positive. The bispectrum can have a positive sign if either all three $\Delta_{21}$ are positive or only one $\Delta_{21}$ is positive. $B(k_1 \sim 0.2 \mp)$ is positive only at the {\em early} stage for almost entire $n-\cos{\theta}$ space, except in the limit and vicinity of the squeezed $k$-triangles. Further, the $B(k_1 \approx 0.7 \mp)$ is positive within the $\cos{\theta}$ range $0.5 \lesssim \cos{\theta} \lesssim 0.85$ for the entire CD except the \vl stage. The bispectra become positive at the {\em very late} stage for $k_1 \gtrsim 1 \mp$ in the entire $n-\cos{\theta}$ space. This is the stage when signal is in emission regime and the matter density fluctuations dominate over the other two constituents of the signal. This results is also in good agreement with the findings of \citet{majumdar20} for the early stages of reionization.

Next, we discuss the sign reversal (change) of the bispectra due to the cosmic evolution of the signal as well as due to the variations in the $k_1$ mode. In the earlier studies \citep{majumdar18, majumdar20, hutter19, watkinson19}, this phenomenon has been associated with the nature of the non-Gaussianity in the 21-cm signal. We first discuss this as a function of $k_1$ mode for a fixed stage of the CD. Figure \ref{fig:b_model_AB} shows that for our fiducial model at the \ve stage as we move from \ss to \ii $k_1$ modes, a sign change of the bispectra (from negative to positive) is observed within $\cos{\theta}$ range $0.5 \lesssim \cos{\theta} \lesssim 0.9$. As one further moves towards \ll $k_1$ modes the sign again gets reversed within the same $\cos{\theta}$ range. A similar feature is also observed for the \m and the \l stages of CD. At the \ee and the \vl stages the scenario is different compared to the other stages. During the \ee stage, as one moves from \ss to \ll $k_1$ mode the sign of the bispectra change first for the linear $k$-triangles and then for the entire $n-\cos{\theta}$ space (except at the limit and the vicinity of the squeezed $k$-triangles). Further, at the \vl stage as we go towards the \ll $k_1$ mode, the bispectra change its sign first for linear $k$-triangles and then for all $k$-triangles. This is due to the fact that as we go towards larger $k_1$ modes at this stage, the matter density fluctuations dominates. This has been identified as the source of the positive sign of the bispectrum.


We next focus on the sign reversal of the bispectra with redshift (i.e. stages of the CD, see Figures \ref{fig:b_model_AB} and \ref{fig:BS_z} for Model-A). During a major portion (i.e., from \ve to \l stages) of the CD when the $T_{\rm S}$ fluctuations dominate over the matter density and the neutral fraction fluctuations, the sign reversal of the bispectra with $z$ is observed only for \ss $k_1$-triangles. For this $k_1$-triangles the sign change takes place twice, first in the transition from the \ve (panel at first column and first row in Figure \ref{fig:b_model_AB}) to the \ee (panel at first column and second row) stage and next in the transition from the \ee to the \m (panel at first column and third row) stage. This happens in almost entire $n$--$\cos{\theta}$ space. The reason behind these sign reversals can be understood via a close investigation of the brightness temperature maps (see first column of Figure \ref{fig:tb_map}). At the \ee stage the Ly$\alpha$ coupling is still dominant over the X-ray heating. However, we see that the heated regions are more numerous (but very small in size) at this stage (compared to the \ve stage) and are distributed following the source locations (i.e. collapsed halos, purple in colour in these maps). As we know that the bispectrum can be positive if either all three $\Delta_{21}$ are positive or only one $\Delta_{21}$ is positive. The maps in Figure \ref{fig:tb_map} suggests that the second reason is most probable during this period, i.e. in most of the cases two length scales (corresponding $k$ modes) belongs to the Ly$\alpha$ coupled regions (giving rise to two negative $\Delta_{21}$s) and one length scale falls in the heated regions (corresponding to one positive $\Delta_{21}$) resulting in the observed positive sign of the bispectra. By the \m stage the Ly$\alpha$ coupled regions still dominate the brightness temperature fluctuations but heated regions are larger in size now. Following same argument now in most of the cases two $\Delta_{21}$s are positive (as they belong to the heated regions) and one $\Delta_{21}$ is negative (belonging the Ly$\alpha$ coupled regions) in their contribution to the bispectrum estimator, resulting in a negative bispectrum.

Finally, we observe one more sign reversal of the bispectra in \ii and \ll $k_1$-triangles and for almost the entire $n$--$\cos{\theta}$ space, during the transition from the \l to the \vl stages. This sign reversal can be understood as follows: At the \l stage the distribution of the heated regions dominates the brightness temperature fluctuations, however there are still a significant amount of Ly$\alpha$ coupled absorption regions present in the IGM (see the panel corresponding to $z = 11.09$ for Model-A in Figure \ref{fig:tb_map}). This makes the bispectrum negative at this stage. By the \vl stage most of the IGM is heated (i.e. $T_{\rm S}$ approaches its saturation limits) and there are some ionized regions present in the IGM (following the distribution of the sources, see the bottom panel for Model-A in Figure \ref{fig:tb_map}). This effectively constitutes the very early stages of the EoR, when the bispectrum is expected to be positive \citep{majumdar18,majumdar20, hutter19}. This particular sign reversal of the bispectrum in \ii and \ll $k_1$ modes can be used as a confirmative test of the saturation of the heating of the IGM and the starting point of a significant global reionization.

Note that in Figure \ref{fig:BS_z} the bispectrum for equilateral $k$ triangle for Model-B shows a sudden change in sign around $z\sim 12$ for large $k_1$ triangles. A similar sign change of this kind is not observed in Model-A in the same or any earlier or later redshifts. Other than this feature, the redshift evolution of the equilateral $k$ triangle bispectrum is quite consistent in both models. Presently, we do not have any obvious interpretation for this feature. In future, we would like to study this in detail by generating many statistically independent realizations of the signal. It will help us to identify if this is caused due to the sample variance or not.  

To draw a comparison between the redshift evolution of the bispectrum and the popular spherically averaged 21-cm power spectrum, we show this power spectrum for three representative $k$ modes in Figure \ref{fig:pk}. A comparison between Figure \ref{fig:BS_z} and \ref{fig:pk} shows that at the \ve stage, both the power spectra and bispectra start with a small magnitude irrespective of the $k$ mode, for both source models.
The magnitude of both of these statistics then increases with decreasing redshift and reach their respective maximum value in the redshift range $z \sim 12-13$. Both of them then gradually decrease their magnitude during the later stages of the cosmic dawn. This is consistent with the evolution of the fluctuations in the 21-cm field that we have discussed earlier. However, apart from the magnitude evolution the bispectrum also shows significant evolution in the sign the details of which depend on the shape of the $k$-triangle and the magnitude of the $k_1$ mode involved. The evolution of the bispectrum thus has the potential to provide us with more insight about the IGM physics compared to what can be derived from the power spectrum.
 

\subsection{Effect of the source models on the bispectrum}

In this section, we compare the redshift space 21-cm bispectrum for two different source models. The details of these two models are tabulated in Table \ref{table1}. We first compare the source models through their 21-cm bispectra for a few specific $k$-triangle shapes, e.g., equilateral, squeezed, L-isosceles and linear $k$-triangles, and then generalize this comparison for all unique triangle shapes in the $n$--$\cos{\theta}$ space. 

First  we check the impact of source models on the shape of the bispectra. A visual inspection of of the evolution of bispectra for equilateral and squeezed $k$-triangles as a function of $z$, shown in Figure \ref{fig:BS_z}, reveals that for a specific $k_1$ mode for both the models the shapes of the bispectra are same. The main features of the shapes of the bispectra in two models only differ by the redshift values at which they appear. Therefore, in Figure \ref{fig:BS_z} the bispectra from two models appear to be just shifted with respect to each other. A similar feature can also be seen in the power spectra (see Figure \ref{fig:pk}).

Next, we compare the bispectra of both the models for L-isosceles and linear $k$-triangles. Figure \ref{fig:line_bispec_all_model} shows the bispectra for Models A and B at five different stages of the CD and for \ss and \ll $k_1$ modes. The first two  and the last two columns represent bispectra for the L-isosceles and linear $k$-triangles, respectively. We first compare the magnitude and sign of the bispectra for both source models at different stages of the CD. An important observation from this figure is that the shape of the bispectra does not depend on the source model, rather it only depends on the stage of evolution of the CD and the specific $k_1$-triangle under consideration. Further, at the \ve stage the magnitude of the bispectra are slightly larger for Model-B than that for Model-A irrespective of the $k_1$ modes and the triangle shapes. Additionally, at this stage the sign of the bispectra are negative in the entire $n$ (for the linear $k$-triangles) and $\cos{\theta}$ (for the L-isosceles $k$-triangles) range. 

At the \ee stages and for small $k_1$ mode for both the L-isosceles and the linear $k$-triangles the magnitude of the bispectra for Model-A are larger than Model-B only when the sign of the bispectra are positive. A reverse phenomenon is observed when the sign of the bispectra are negative. Furthermore, at the \ee stage although the shape of the bispectra for both models are same, the sign of the bispectra for small $k_1$ triangles get reverse for small values of $\cos{\theta}$ in Model-B. On the other hand, at this stage for \ll $k_1$ triangles the magnitude, the sign and the shape of the bispectra for both the models are same.

We next observe that at the \m stage of the CD for the \ss L-isosceles and the linear $k_1$-triangles the sign of the bispectra for Model-A are negative and for Model-B are positive while their magnitudes are same. Further, at this stage for \ll L-isosceles and linear $k_1$-triangles the magnitude of the bispectra are larger in Model-B, but, they have same signs and the shapes as Model-A.

At the \l stage of the CD the magnitude of the bispectra for Model-B are larger than that for Model-A irrespective of the $k_1$ mode and $k$-triangle shapes. The sign of the bispectra are same for both models irrespective of the $k_1$ mode $k$-triangle shapes.

Finally, at \vl stage we observe a large difference between magnitude of the bispectra for Model-A and B at \ss $k_1$ mode. It is large by $\sim 2$ orders of magnitude for Model-B as compared to Model-A. This is due to the fact that for Model-A, by the \vl stage most of the IGM is heated (i.e. $T_{\rm S}$ approaches its saturation limits) however, for Model-B there are still some leftover unheated Ly$\alpha$ coupled regions (see bottom panel of Figure \ref{fig:tb_map}). This causes a larger fluctuation in the signal for Model-B compared to Model-A at these redshifts, leading to a higher magnitude of the bispectra. This feature implies that due to the presence of only HMXB type sources (i.e. Model-B) the IGM will get heated completely at a later time as compared to the scenario when there are only mini-QSO (Model-A) type sources. Further, at this stage for small $k_1$ triangles the sign and the shape of the bispectra are same in both models. Lastly, for large $k_1$-triangles the sign of the bispectra are positive in both models and the magnitude of the bispectra in Model-A is slightly higher than that in Model-B. Hence we can conclude that the different source models do not significantly affect the shape of the bispectra but, they change its magnitude (significantly) and sign depending on the stages of the CD and $k$-triangle shape for which the bispectra are being estimated.

Next, we generalize our discussion for all unique $k$-triangles in the $n$--$\cos{\theta}$ space. In the last three columns of Figure \ref{fig:b_model_AB} we show the bispectra for all unique $k$-triangles for Model-B at five different stages (at the same redshifts as for Model-A shown in Figure \ref{fig:b_model_AB}) of the CD and at three different $k_1$ modes. We now discuss the impact of the different sources on the evolution of the bispectra through a comparison between Model-A and Model-B. We first compare their redshift evolution at \ss $k_1$ mode. For triangles with \ss $k_1$ mode, as one make transition from the \ve stage to the \ee stage, the sign reversal of the bispectra for Model-B is observed only in the limit and the vicinity of the S-isosceles $k$-triangles. On the other hand for Model-A it is observed almost in the entire $n$--$\cos{\theta}$ space. For Model-B a further transition from \ee to \m stage at \ss $k_1$ mode makes the bispectra sign positive almost in the entire $n$--$\cos{\theta}$ space. However, during the same transition in Model-A the bispectra sign changes from positive to negative in for all triangle shapes. In the next obvious transition, from \m to \l stage, for Model-B we observe a complete sign change (positive to negative) of the bispectra for in the entire $n$--$\cos{\theta}$ space. On the contrary, for Model-A no further sign change is observed. Lastly, in the transition from \l to \vl stage, for Model-B no sign reversal is observed (similar to Model-A), rather, a decrease in the magnitude of the bispectra is observed in the entire $n$--$\cos{\theta}$ space. This decrements in the magnitude for Model-B is small as compared to what is observed for Model-A in this regime. In fact around \vl stage, for Model-B, in the entire $n$--$\cos{\theta}$ space, the magnitude of the bispectra is larger by $\sim 2$ orders of magnitude compared the same for Model-A. This is due to the same physical reason as discussed earlier in this section for L-isosceles and linear $k$-triangles. Hence triangles of all shapes, in the entire $n$--$\cos{\theta}$ space, carries the signature of the late time saturation of IGM heating by the HMXB type sources as compared to the scenario for mini-QSO type sources.

We observe that the two source models have similar impact on the sign of the bispectra (and its change with redshift) from the \ve to the \vl stages of the CD for the \ii and \ll $k_1$ modes. The only difference in their impact shows up through the magnitude of the bispectra for these $k_1$ modes (similar to the discussion earlier in this section for L-isosceles and linear $k$-triangles for the large $k_1$ mode).

\begin{figure*}
  \includegraphics[width=1\textwidth,angle=0]{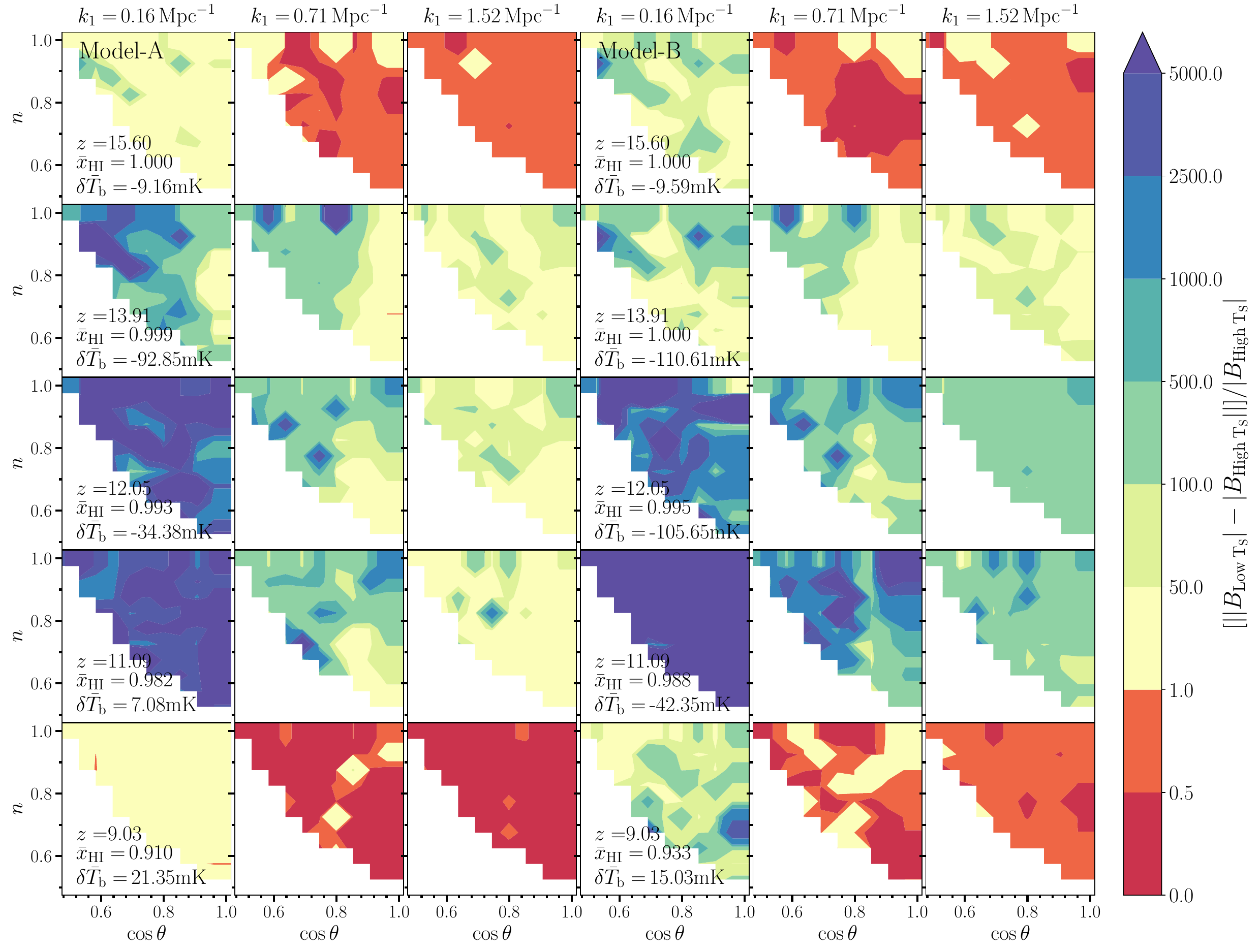}
  \caption{The quantity $\mathfrak{D}(k_1,\, n,\, \cos{\theta}) = [||B_{\rm Low \:T_S}|-|B_{\rm High\: T_ S}||]/|B_{\rm High\: T_ S}|$,  estimating the effect of the spin temperature fluctuations on the bispectrum magnitude for Model-A (left three columns) and Model-B (right three columns) at five different stages of the CD and for three different $k_1$ modes.}
  \label{fig:mod_diff}
\end{figure*}
\begin{figure*}
  \includegraphics[width=1\textwidth,angle=0]{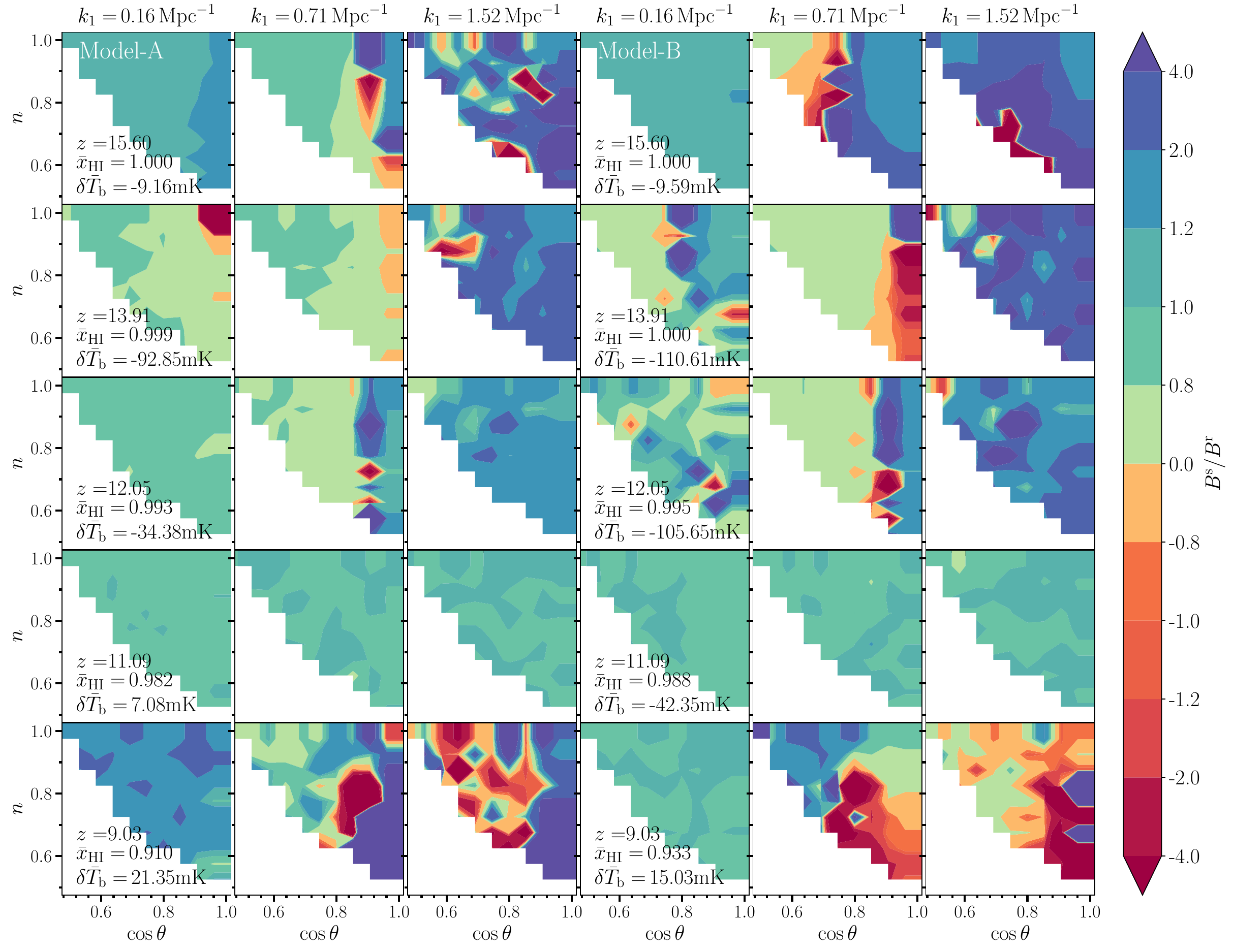}
  \caption{The ratio $B^{\rm s}/B^{\rm r}$, quantifying the impact of the RSD on the CD 21-cm bispectra for all unique triangle configurations of Model-A and Model-B at five different stages of the CD and for three different $k_1$ modes.}
  \label{fig:bs_to_br}
\end{figure*}
\subsection{Impact of the spin temperature fluctuations on the bispectrum}
\label{sec:impact_of_Ts}
The spin temperature fluctuation is the component (the other two components being the matter density and the neutral fraction) that has maximum contribution in determining the nature and magnitude of the 21-cm brightness temperature fluctuations during the CD. Consequently one would expect that it will have a significant impact on the bispectrum of the signal as well. One of the main aim of this article is to study the impact of the spin temperature fluctuations on the bispectrum statistic. To quantify this impact, we compute the quantity $\mathfrak{D} = [||B_{\rm Low \:T_S}|-|B_{\rm High\: T_ S}||]/|B_{\rm High\: T_ S}|$  (see Figure \ref{fig:mod_diff}) where, $B_{\rm Low \,T_ S}$ is the bispectrum values when fluctuations in the $T_{\rm S}$ has been taken under consideration. On the other hand, $B_{\rm High \,T_S}$ is the bispectrum values at the limit when $T_{\rm S}$ reaches saturation i.e., $T_{\rm S} = T_{\rm K} \gg \TCMB$. Thus this quantity $\mathfrak{D}$ will also be a function of $k_1,\, n,\, \cos{\theta}$, just as the bispectrum itself. A high value of $\mathfrak{D}$ corresponds to a large impact of the spin temperature fluctuations on the bispectra and vice versa. Figure \ref{fig:mod_diff} shows $\mathfrak{D}(k_1,\, n,\, \cos{\theta})$ for both Models A and B, at three different $k_1$ modes and five different stages of the CD.

We first note the $k_1$ dependence of $\mathfrak{D}$ from Figure \ref{fig:mod_diff}. We find that irrespective of the source models and stage of the CD and for almost all unique triangle shapes, $\mathfrak{D}$ decreases monotonically as one varies $k_1$ modes from \ss to \ll. $\mathfrak{D}$ has largest magnitude for \ss $k_1$-triangles and smallest magnitude for \ll $k_1$-triangles. The reason for this behaviour of $\mathfrak{D}$ can be understood in the following manner: the smaller $k$-modes will be sensitive to the signal coming from the larger length scales. The signal fluctuations at large scales (i.e. smaller $k$-modes) is significantly influenced by the presence of large number of absorption and emission regions (having sizes smaller than the large length scale under consideration). Thus one would expect the magnitude of bispectra to decline monotonically from  \ss to \ll $k_1$-triangles.

We next discuss the redshift evolution of $\mathfrak{D}$. Irrespective of the source model and $k_1$ mode, one can clearly see a general trend in the evolution of $\mathfrak{D}$. The $\mathfrak{D}$ is small at the \ve stage. It then gradually increases and reaches a maxima around the \m stage for Model-A and around the \l stage for Model-B. After the \m stage it gradually decreases until the \vl stage. One can see a clear one-to-one correspondence between the evolution of $\mathfrak{D}$ with redshift and the evolution of the amplitude of fluctuations in the signal with redshift (see Figure \ref{fig:tb_map}).

\subsection{Impact of the RSD on the CD bispectrum}
We next focus on quantifying the impact of RSD on the 21-cm bispectrum from the CD. As per our knowledge this is the first attempt to quantify the effect of RSD on the CD 21-cm bispectrum using simulated signals. To quantify the impact of RSD, we show the ratio of the spherically averaged bispectrum in redshift space ($B^{\rm s}$) to its real space counterpart ($B^{\rm r}$), i.e., $B^{\rm s}/B^{\rm r}$ in Figure \ref{fig:bs_to_br}. The figure shows redshift evolution of the $B^{\rm s}/B^{\rm r}$ for both models and for three different $k_1$ modes. The RSD affects both magnitude and sign of the bispectra. The values of the $B^{\rm s}/B^{\rm r}$ in this figure can be divide into three groups: $B^{\rm s}/B^{\rm r} \approx 1 $, implying that the impact of the RSD is negligible, $B^{\rm s}/B^{\rm r} > 1$ implying that the RSD enhances the magnitude of the bispectra and $0< B^{\rm s}/B^{\rm r} < 1$, implying that RSD reduces the magnitude of the bispectra. Additionally, if  $B^{\rm s}/B^{\rm r} < 0$, it implies that the RSD changes the sign of the bispectra. We label this sign change as {\em sign difference} between the real and redshift space bispectrum.

Starting at the \ve stage, we observe that the RSD enhances the magnitude of the bispectra (by $\lesssim 100\%$ for Model-A and  $\lesssim 10\%$ for Model-B) for \ss $k_1$ modes, in the entire $n$--$\cos{\theta}$ space. For \ii and \ll $k_1$ modes the impact is even larger. The RSD either changes (increases or decreases) the magnitude of the bispectra by more than $20\%$ or/and introduces {\em sign difference} for these $k$-triangles. As the Ly$\alpha$ coupling and X-ray heating of the IGM progresses further, for different source models RSD impacts differently depending on the stages of the CD and $k_1$ modes involved.

In the \ee stages for \ss $k_1$ mode, the {\em sign difference} arises in the limit and vicinity of the squeezed $k$-triangles in Model-A, however, for Model-B no such {\em sign difference} is observed. Furthermore, for other $k$-triangles $n$--$\cos{\theta}$ space (for both models), the RSD decreases the bispectrum magnitude by more than $20\%$. For \ii and \ll $k_1$-triangles in the entire $n$--$\cos{\theta}$ space, RSD has the largest impact on both magnitude (changes as high as $\sim 400\%$) and sign, irrespective of the source models.

While transitioning form the \ee to the \m stage, the impact of RSD decreases until the \l stage is reached (irrespective of $k_1$ modes and source types). Hence, the RSD has minimum impact around the \l stages for both source models and for all $k_1$-triangles. A further transition from the \l to the \vl stage allows the impact of RSD on the bispectra to increase.

Finally, we discuss the impact of the RSD around the \vl stage. For \ss $k_1$ mode, it affects only the magnitude of the bispectra for Model-A, enhancing it by more than $\sim 20\%$. On the other hand, for \ii and \ll $k_1$ modes, the RSD changes both the magnitude and the sign of the bispectra irrespective of source model. 

\subsection{Interpretation of the CD redshift space bispectra using the linear theory of the RSD}

To interpret the features observed in the spherically averaged redshift space bispectra, we opt for the linear model for the differential brightness temperature in redshift space while considering the effect of the $\TS$ fluctuations \citep{bharadwaj04, bharadwaj05, pritchard08}. Under this linear model one can express the spherically averaged CD redshift space bispectra as a correction to the real space bispectra as shown in Section 3 of Paper I. In Paper I (Section 5.2.2 and Appendix A) we have demonstrated that the correction term that has maximum impact on the redshift space bispectra is the $B_{\mu^2-{\rm RC}}$. In case of the CD 21-cm signal bispectra that we consider here, we find the same to be true (demonstrative figures not shown here and we request the interested readers to refer to Paper I).

\comment{
To interpret the features observed in the spherically averaged redshift space bispectra, we use Equation \eqref{eq:b_qlin} as our model for the redshift space bispectra, which is derived using the linear theory of redshift space distortions in Section \ref{sec:B_rec}. We label our model bispectra as the {\em reconstructed} redshift space bispectra ($B^{\rm s, rec}$). Equation \eqref{eq:b_qlin} has total eight terms in its right hand side, of which the first term is the auto-bispectrum of the 21-cm field, which is nothing but the real space bispectrum ($B^{\rm r}$). The other seven terms are the Redshift space Correction (RC) terms or the combination of different cross-bispectra terms of the 21-cm and matter density fields weighted by the different degrees of $\mu$ as their coefficients. We estimate the {\em reconstructed} bispectra ($B^{\rm s, rec}$) from the simulated real space 21-cm and the matter density fields. Further, we label the redshift space bispectra ($B^{\rm s}$) estimated from the simulated redshift space 21-cm field as {\em original} bispectra. We next test the validity of this model at different stages of the CD and for different $k_1$ modes by estimating the ratio of the {\em reconstructed} bispectra to the {\em original} bispectra i.e. $B^{\rm s, rec}/B^{\rm s}$. Figure \ref{fig:Brec_Borigin} shows this ratio, $B^{\rm s, rec}/B^{\rm s}$, for our fiducial source model at four different stages (\ve, \ee, \m and \vl) of the CD and for \ss and \ll $k_1$ modes. This figure shows that the linear model of the RSD can reconstruct the redshift space bispectra very well when compared to the {\em original} redshift space bispectra for \ss $k_1$-triangles. The $B^{\rm s, rec}$ deviates from the $B^{\rm s}$ by a rather small amount ($\sim 5$--$20\%$) for a large period during the CD. This model can reproduce both the magnitude and sign of the bispectra at various stages of the CD. At the \ve and the \vl stages the $B^{\rm s, rec}$ deviates significantly from the $B^{\rm s}$ in magnitude, however, it is still able to reproduce the sign and shape of the signal bispectra correctly. Further, for \ll $k_1$-triangles $B^{\rm s, rec}$ deviates significantly from the $B^{\rm s}$ (more than $\sim 20\%$). This is possibly caused due to the presence of small scale features e.g. the Finger-of-God effects \citep{jackson72} etc. in the signal, which the linear theory of redshift space distortions is not expected to model very well. 

As discussed in Section \ref{sec:B_rec}, we divide the seven RC terms into three groups and label them as $B_{\mu^2-{\rm RC}}$, $B_{\mu^4-{\rm RC}}$ and $B_{\mu^6-{\rm RC}}$. To quantify each of these three groups relative contribution in shaping the $B^{\rm s, rec}$, we estimate the following three ratios:    $B_{\mu^2-{\rm RC}}/B^{\rm r}$, $B_{\mu^4-{\rm RC}}/B^{\rm r}$ and $B_{\mu^6-{\rm RC}}/B^{\rm r}$. From these ratios we find that $B_{\mu^4-{\rm RC}}$ and $B_{\mu^6-{\rm RC}}$ has less than $10\%$ contribution (in units of $B^{\rm r}$) in shaping the $B^{\rm s, rec}$. The component that drives the deviations in $B^{\rm s, rec}$ when compared to $B^{\rm r}$ is $B_{\mu^2-{\rm RC}}${\footnote{Note that a similar behaviour of $B_{\mu^2-{\rm RC}}$ was observed in our earlier work \citep{majumdar20} where we try to model the redshift space 21-cm bispectra during the reionization.}}. In Figure \ref{fig:B2_and_4_to_Br} we show its relative contributions in $B^{\rm s, rec}$. We next compare Figure \ref{fig:B2_and_4_to_Br} with Figure \ref{fig:bs_to_br} ($B^{\rm s}/B^{\rm r}$) to get an idea how the magnitude and sign of the redshift space bispectra is affected by the $B_{\mu^2-{\rm RC}}$ term.  

At the \ve stages of the CD and for small $k_1$-triangles the ratio $B_{\mu^2-{\rm RC}}/B^{\rm r}$ is positive and has its magnitude in the range $\sim 0.5$--$1.0$ (Figure \ref{fig:B2_and_4_to_Br}). This implies that $B_{\mu^2-{\rm RC}}$ contributes in the redshift space signal bispectra with the same sign as $B^{\rm r}$, which further implies that it will help in enhancing the magnitude of the $B^{\rm s}$ with respect to $B^{\rm r}$. We see a confirmation of this prediction in Figure \ref{fig:bs_to_br} which shows the ratio $B^{\rm s}/B^{\rm r}$ ranging between $\sim 1$--$2$ at the \ve stages for small $k_1$-triangles.

Around the \ee and the \m stages, however, the $B_{\mu^2-{\rm RC}}/B^{\rm r}$ has a negative sign. This implies that $B_{\mu^2-{\rm RC}}$ contributes in the redshift space signal bispectra with the opposite sign of $B^{\rm r}$, which further implies that it will help in reducing the magnitude of the $B^{\rm s}$ with respect to $B^{\rm r}$. This makes $B^{\rm s}/B^{\rm r} < 1$ in most of the $n-\cos{\theta}$ space. This is also the cause of the {\em sign reversal} at the \ee stage in the limit and vicinity of the squeezed $k$-triangles. Lastly, around the \vl stages the $B_{\mu^2-{\rm RC}}$ has its largest contribution with the same sign as $B^{\rm r}$. Thus it enhances the magnitude of the $B^{\rm s}$ with respect to $B^{\rm r}$ significantly.
}

%% file: summary.tex
\section{Summary and Discussions}
In this paper, we have presented a comprehensive view of the non-Gaussianity in the 21-cm signal from the Cosmic Dawn (CD) through the signal bispectra estimated for all unique $k$-triangles. All of the earlier works in this line have considered only a few specific types of $k$-triangles. We explore the $k$-triangles, ${\bf k}_1 +{\bf k}_2 +{\bf k}_3 =0$, using $k_1$, $n={k_2}/{k_1}$ and $\cos{\theta} = -({{\mathbf k}_1\cdot {\mathbf k}_2})/({k_1 k_2})$ quantities. Additionally, for the first time, we also quantify the impact the redshift space distortions as well as the impact of the spin temperature fluctuations on the CD 21-cm bispectra for all unique $k$-triangles. The characteristics of the heating and ionizing  sources in the early universe may have a significant impact on the signal bispectrum. To understand the impact of the source models on the bispectra, we have considered two different types of sources in this work, they are mini-QSOs (Model-A) and HMXBs (Model-B). Note that the entire analysis presented in this paper is based on the redshift space 21-cm signal from the cosmic dawn. The findings of our analysis can be summarized as follows:

\begin{itemize}
    \item The magnitude of the CD 21-cm bispectra in the entire $n$--$\cos{\theta}$ space initially increases with decreasing redshift irrespective of the $k_1$ modes and source model. This is due to the increasing amplitude of fluctuations in the signal due to the simultaneous Ly$\alpha$ coupling and X-ray heating of the IGM. The bispectrum reaches its maxima around the stage when the signal fluctuations are also at their maximum (and the Ly$\alpha$ coupling reaches its saturation). After that the bispectrum magnitude decreases with decreasing redshift as further heating of the IGM decreases the amplitude of fluctuations in the signal. Finally, the magnitude of the bispectrum is minimum when most of the IGM is heated.
    
    \item The redshift evolution of the magnitude of the bispectra for both source models show similar trends, however the redshifts at which the maxima and minima of the amplitude appears depend on the nature of the sources of heating and ionization. For mini-QSO type sources maxima and minima of the bispectra magnitude appear around $z \sim 12$ and $z \sim 9$, respectively. Whereas for HMXB type sources they appear around $z \sim 11$ and $z \lesssim 9$. Bispectra for triangles of almost all unique shapes show this behaviour. This is a signature of the rather late saturation of IGM heating in case of the HMXB type sources as compared to the mini-QSO type sources.
    
    \item The sign of the CD 21-cm bispectra is an important feature of this statistic. For a given stage of the CD and a specific length scale, the sign of the bispectra effectively probes which among the three physical processes, namely heating, Ly$\alpha$ coupling and reionization, is the major cause of non-Gaussianity in the signal. 
    
    \item The sign reversal of the bispectra is an important phenomenon associated with the nature of the non-Gaussianity in the signal. In scenarios when both Ly$\alpha$ coupling and X-ray heating are underway simultaneously (in a competitive manner), we observe two sign reversals in bispectra for \ss $k_1$-triangles (irrespective of the source models). The first one is from negative to positive and the second one from positive to negative (see first and fourth columns in Figure \ref{fig:b_model_AB} for Model-A and Model-B respectively). Via a visual inspection of the simulated 21-cm maps we associate the first sign reversal with the emergence of the numerous but very small sized heated regions in the IGM while the Ly$\alpha$ coupling still have dominating contribution to the signal fluctuations  and thus to its non-Gaussianity. The second sign reversal is associated with the phase when the Ly$\alpha$ coupling in the entire IGM is complete. Such distinct signature of these two unique phases of the CD is not visible in any other signal statistics e.g. the redshift evolution of the mean global signal or the power spectrum. Thus the sign reversals of bispectrum have the possibility of providing us a unique and independent way of constraining the CD history. To check the robustness of this phenomenon we need to study it for a variety of source models and cosmic dawn scenarios which we plan to take up in a follow up work.

    \item The spin temperature fluctuations have significant impact on the signal bispectra. Independent of the source models, the impact is maximum on the bispectra for triangles with \ss $k_1$ modes. Further, for a specific $k$-triangle the impact of $T_{\rm S}$ fluctuations on the signal bispectra initially increases with decreasing $z$ due to the simultaneous Ly$\alpha$ coupling and X-ray heating of the IGM. The impact $T_{\rm S}$ fluctuations is largest around the stage when the Ly$\alpha$ coupling reaches saturation. After this the impact of the $T_{\rm S}$ fluctuations gradually decreases as X-ray heating takes up the dominant role in shaping the signal and overall amplitude of fluctuations in the signal brightness temperature also goes down.  
    
    \item The redshift space distortions have a significant impact on both the magnitude and the sign of the bispectra. The level of this impact depends on the specific stage of the CD and $k$-triangle shape and size. At any given stage of the CD, its maximum impact is observed at bispectra for the large $k_1$-triangles. For most of scenarios RSD changes the magnitude of the bispectra by more than $\sim 20\%$.
    
    \item We find that a model bispectra based on the linear theory of the RSD provides a good interpretation the simulated bispectra observed for triangles with small $k_1$ modes. Further, among the three groups of correction terms that contributes in the model bispectra, the $B_{{\mu}^2-{\rm RC}}$ dominates in shaping the CD 21-cm redshift space bispectra amplitude and sign.
\end{itemize} 

The analysis of the CD 21-cm redshift space bispectra in this article though of very comprehensive in nature but is limited to only two source models, both having same model parameters like $M_{\rm halo,\, min}$, $f_{\star}$, $f_{\rm esc}$ etc. To properly understand the impact of the source models on the signal bispectra, we will need to study it for a much wider variety of sources, which we plan to explore in a follow up work.

The CD scenarios considered here (for both source models) have both the Ly$\alpha$ coupling and the X-ray heating of the IGM running in parallel from the \ve stages of the CD. As discussed before this may have a significant impact on the bispectra amplitude, sign and sign reversal. We understand the robustness of our results we need to also study the scenarios where first the Ly$\alpha$ coupling of the IGM reaches saturation and then X-ray heating starts to contribute significantly. This also we plan to explore in a future work.

This work is focused on quantifying the impact of the RSD on the spherically averaged bispectra. For a more complete understanding of the anisotropy in the signal bispectra due to redshift space distortions, one would need to decompose the bispectra into a vector space of the orthonormal basis as suggested by \citet{bharadwaj20}.

Furthermore, in the present analysis we have not considered another important LoS effect which is always present in an actual observation, the light cone effect \citep{datta12, datta14, mondal18}. This LoS anisotropy in the signal arises due to the time evolution of the cosmic 21-cm signal. 

Lastly, in this work we have not considered the effects caused by incomplete foreground subtraction. It is essential to use optimal foreground removal techniques for a reliable estimation of the signal bispectra from the observed data \citep{watkinson20, 2021MNRAS.500.2264H}. Incomplete foreground removal may result in a wrong interpretation of the signal bispectra. Furthermore, we have also not discussed the detectability of the 21-cm bispectra for any present or future telescopes, taking into account the presence of thermal noise \citep{yoshiura14} and other systematic uncertainties. We plan to address these issues in our followup work. 
\label{sec:summary}

%% file: acknow.tex
\section*{Acknowledgements}
SM and GM acknowledge financial support from the ASEM-DUO India 2020 fellowship. JRP and SM acknowledge financial support from the European Research Council under ERC grant number 638743-FIRSTDAWN. RM is supported by the Wenner-Gren Postdoctoral Fellowship. RG furthermore acknowledge support by the Israel Science Foundation (grant no. 255/18). ITI is supported by the Science and Technology Facilities Council [grant numbers ST/I000976/1 and ST/T000473/1] and the Southeast Physics Network (SEP-Net). GM is supported in part by Swedish Research Council grant 2016-03581. Some of the numerical results were obtained using the PRACE Research Infrastructure resources Curie based at the Très Grand Centre de Calcul (TGCC) operated by CEA near Paris, France and Marenostrum based in the Barcelona Supercomputing Center, Spain. Time on these resources was awarded by PRACE under PRACE4LOFAR grants 2012061089 and 2014102339 as well as under the Multi-scale Reionization grants 2014102281 and 2015122822. The major part of the radiative transfer simulations (excluding the N-body simulations and halo identification) and bispectrum estimation were done using the computing resources available to the Cosmology with Statistical Inference (CSI) research group at IIT Indore.